# The brain versus AI: World-model-based versatile circuit computation underlying diverse functions in the neocortex and cerebellum

Shogo Ohmae[1,*] & Keiko Ohmae[1]

[1] Chinese Institute for Brain Research (CIBR), Beijing, China

* Correspondence to: shogo@cibr.ac.cn

## Abstract

AI's significant recent advances using general-purpose circuit computations offer a potential window into how the neocortex and cerebellum of the brain are able to achieve a diverse range of functions across sensory, cognitive, and motor domains, despite their uniform circuit structures. However, comparing the brain and AI is challenging unless clear similarities exist broadly, and past reviews have been limited to comparison of brain-inspired vision AI and the visual neocortex. Here, to enable comparisons across diverse functional domains, we subdivide circuit computation into three elements — circuit structure, input/outputs, and the learning algorithm — and evaluate the similarities for each element. With this novel approach, we identify wide-ranging similarities and convergent evolution in the brain and AI, providing new insights into key concepts in neuroscience. Furthermore, inspired by processing mechanisms of AI, we propose a new theory that integrates established neuroscience theories, particularly the theories of internal models and the mirror neuron system. Both the neocortex and cerebellum predict future world events from past information and learn from prediction errors, thereby acquiring models of the world. These models enable three core processes: (1) Prediction — generating future information, (2) Understanding — interpreting the external world via compressed and abstracted sensory information, and (3) Generation — repurposing the future-information generation mechanism to produce other types of outputs. The universal application of these processes underlies the ability of the neocortex and cerebellum to accomplish diverse functions with uniform circuits. Our systematic approach, insights, and theory promise groundbreaking advances in understanding the brain.

# Introduction

AI development and brain research have a long history of mutual influence. In light of recent advancements in AI, there is a high expectation that studying the brain with reference to brain-like AI will lead to new theories and concepts in neuroscience [1-10]. In the past, AIs generally had a single function, but since 2018, general-purpose AIs have emerged that are capable of performing multiple functions with a single circuit. By comparing the brain with these general-purpose AIs, we may be able to unravel the enduring mystery of how the neocortex and cerebellum are each able to execute a wide variety of functions across multiple domains, despite their relatively uniform circuit structures and local-circuit computations [2,11-14]. To address the mystery of the brain's universal local-circuit level computations (hereafter, circuit computations), a comprehensive comparison of the brain and AI across various functional domains is essential. Such comparison is straightforward when there is a high degree of similarity in most aspects of the circuit computations; however, when there is only partial similarity, it is challenging to assess which aspects of the circuit computations are similar and to what extent, and to avoid merely listing fragmentary similarities. In fact, previous reviews comparing the brain and AI have been heavily biased toward visual neocortex and brain-inspired vision AI because of their high degree of similarity, while failing to address representative neuroscience theories or historically successful AI circuits in other functional domains [1,3-7,15]. A comparison that focuses solely on visual processing is insufficient to satisfy the needs of AI experts seeking new insights from the brain, neuroscientists looking for novel perspectives from AI, or, more broadly, anyone curious about the secrets of human intelligence.

To provide a comprehensive and structured comparison, we decided to break down the circuit-computation mechanisms into three elements and analyze the similarities between the brain and AI for each element. (i) Circuit structure: In the brain, specific constraints on circuit architectures are imposed by the particular neuronal subtypes, the anatomical neuronal connections, and the learning sites at the connections. By contrast, the design of AI circuits is flexible and can be adjusted according to the task demands. In both cases, the circuit structure is a critical factor that determines the upper limit of the capacity of the circuit computation. For example, the structure of an AI circuit determines its limits, which cannot be surpassed even with large-scale learning [14,16]. (ii) Input/output signals: Information processing in the brain is fundamentally a transformation from input to output, so the input/output signals correspond one-to-one to the function of the circuit. (iii) Circuit learning: Even if the input/output characteristics of circuits are similar, different learning methods (e.g., supervised vs. unsupervised learning) can result in different intermediate processes from one circuit to the next [17-23]. By breaking down the circuit computation into these three elements and evaluating the similarities for each, we are able to comprehensively compare the brain and AI across sensory, cognitive, and motor domains for the first time. This new approach clarifies which aspects of circuit computation contain brain–AI similarities, allowing us to draw new insights from AI and relate these to key neuroscience theories and concepts.

Through our review, we find that modern AI has repeatedly undergone unexpected convergences in information processing with the brain, revealing many similarities beyond those aspects intentionally designed to imitate the brain. Drawing on recent large-scale AI models that have shown particularly interesting convergence with the brain, we propose a new theory integrating existing neuroscience theories (relating to internal models, sensory processing, and the mirror neuron system) to explain the universal circuit computations underlying the diverse functions of the neocortex and cerebellum: The neocortex and cerebellum predict future states of the external world from past world information and learn to minimize prediction errors. Through this process, they develop models of the world, thereby acquiring three types of information processing.

(1) Prediction: Utilizing the world models to generate future information.
(2) Understanding: Utilizing compressed and conceptualized information within the world models to understand the external world (e.g., object recognition, sentence comprehension).
(3) Generation: Repurposing the future-information generation mechanism (i.e., the predictive function) to produce other types of information output (e.g., action planning, language planning, and imitation learning).

Combining these three fundamental information processes and applying them across sensory, cognitive, and motor domains is the key to the powerful, general-purpose information processing shared by the neocortex, cerebellum, and recent large-scale AI.

Through this review, we also seek to organize key AI knowledge that biological neuroscience should consider in this new era of widespread AI, clarify the major questions that neuroscience can leverage AI to address, and propose directions for future neuroscience research. In this way, our review responds to the contemporary needs of neuroscientists as well as provides new insights into key topics in neuroscience, such as attention mechanisms, reward signals, the default mode network, and neurodevelopmental disorders. Traditional approaches to conceptualizing brain functions have overly relied on verbal descriptions, often neglecting aspects that are difficult to verbalize. By referencing AI processing that resembles or imitates the brain, neuroscientists will be able to address these previously overlooked aspects, potentially leading to significant advances in our understanding of the brain.

# 1. Sensory Information Processing

The ultimate goal of sensory information processing is to accurately infer the state of the external world from limited sensory signals through internal "world models" in the neocortex [12,24-35] and in the cerebellum [14,36-43], which are acquired through sensory experience. In vision, for example, the goal is to reconstruct 3D models of objects and space from 2D retinal images [44,45]. In this section, we compare sensory brain circuits with AI circuits to deepen our understanding of circuit computation in sensory information processing in the brain. We start with object recognition in vision, the field where the theoretical understanding of the brain and AI circuits have most closely influenced each other.

## 1.1 Visual Processing

**Deep-layered autoencoders as a circuit computation theory of the visual neocortex**

The visual recognition circuit, located in the cerebral visual neocortex, is primarily trained through unsupervised learning [4,15,21,46-48]. Generally, the brain must learn to perform appropriate information processing without explicit instruction on the correct processing or outputs [15,48]. In the visual system, for instance, humans learn to distinguish animals such as dogs from other objects during childhood. However, during this learning process, our visual system is not explicitly taught the correct boundary between dogs and other visual objects, nor the correct instruction for dog recognition (i.e., the correct label). Learning under such conditions is known as unsupervised learning, and the sensory neocortices, including the visual neocortex, are considered representative unsupervised-learning circuits [21,46]. According to this learning theory of the neocortex, a representative strategy for learning without explicit instruction is unsupervised "autoencoder" learning, where the original input image serves as the target information to learn [15,21]. The circuit is trained such that the input neurons receive a 2D visual image and the downstream restoration neurons reconstruct the original input (Figure 1a). Because the output needs to encode the input, this type of circuit is termed an "autoencoder" in machine learning. When there are fewer neurons in the middle layer than in the input neurons, information needs to be compressed into a lower-dimensional space (Figure 1a, red), resulting in the generation of compressed visual information as a secondary output.

The idea that the visual neocortex can function as an autoencoder has a long history [21]. In fact, when an autoencoder circuit of an artificial neural network (ANN) was trained with numerous natural images, the compressed-representation neurons acquired receptive fields similar to those of the primary visual neocortex. That is, the optimal visual inputs to activate the neurons were local line/stripe patterns, described by Gabor functions, that detect object contours and edges (Figure 1b). Interestingly, the compression in those neurons reflects the features of natural images; when trained with artificial images with non-natural features, the receptive fields of the neurons adapted. This observation aligns with the finding that cats raised in an environment with vertical stripes develop visual neurons with receptive fields that predominantly prefer vertical lines/stripes [49]. This similarity suggests that the receptive fields in primary visual neocortex develop through a process of feature extraction from natural images analogous to an autoencoder.

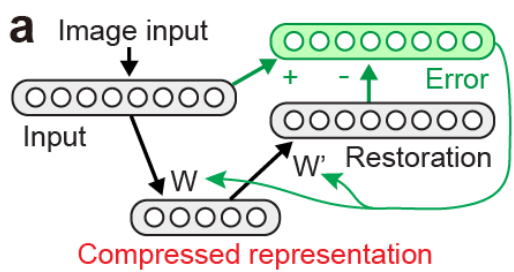

**Figure 1. Deep-layered autoencoders and CNNs as theories of visual information processing in the neocortex.**

**a**, Autoencoder circuit. In a representative autoencoder circuit, the input neurons receive a 2D visual image (e.g., the pixel information of a photograph). The restoration neurons (downstream via disynaptic projections) are trained to encode the same information as the input neurons (In 1996, Olshausen & Field trained the circuit with constraints on synaptic weights, namely L1 regularization, although recent large-scale AI circuits can often be trained without explicit constraints, i.e., by implicit regularization). If the number of downstream neurons is equal to or greater than the number of input neurons, the autoencoder function can be achieved simply by passing the image information along as it is. However, if the downstream has fewer neurons, the input information needs to be compressed into a more compact dimensional space (red; compressed representation neurons). The error neurons calculate the difference between the input neurons and the restoration neurons, and the synaptic weights (W and W') of the circuit are updated to minimize this error.

Circuit structure: This representative autoencoder circuit is a classical three-layer neural network (although various types of circuits can be trained to acquire the autoencoder function). Input/output: The input is a 2D visual image, and the output is the image reconstructed by the restoration neurons. Additionally, the signals from the compressed-representation neurons (red) serve as a secondary but more processed output (i.e., a de facto output). Learning: The synaptic weights (W and W') are updated to minimize the reconstruction error through unsupervised learning. The error signal (generated by the error neurons) is used solely for learning (green) and does not contribute to the outputs of the information processing.

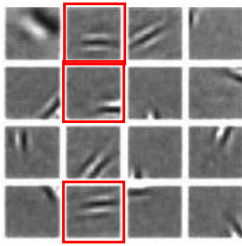

**Figure 1b.** Neurons in the autoencoder circuit in **a** acquired receptive fields with local line/stripe patterns that can be described by Gabor functions, as in the primary visual neocortex. The red frames highlight neurons with receptive fields that prefer horizontal lines/stripes in different local areas of the visual field. Panel **b** adapted from Olshausen & Field (1996) with permission from Springer Nature.

Later, to achieve more advanced information processing to extract more abstract visual features, researchers attempted to create a deep-layered (hierarchical) autoencoder by stacking the data-compression functions of autoencoders. The entire block shown in Figure 1a is referred to here as a layer and is stacked multiple times such that the signal of the compressed-representation neurons is the output to the next layer. Initially, training all layers simultaneously by backpropagation proved challenging. However, in 2006, Hinton and colleagues succeeded in training a deep-layered autoencoder using a method called greedy layer-wise training, in which the deep autoencoder was trained one layer at a time, starting from the lowest layer (that received the raw image input) [50,51]. This learning procedure is analogous to the critical period in the visual neocortices (during which the lower visual neocortices largely determine the characteristics of information processing at an early stage in life) [52].

  Furthermore, in 2012, a group of researchers at Google successfully enabled neurons at the highest layer of a deep autoencoder to acquire specific responses to and recognition of human and cat faces. This was achieved by implementing tolerance sub-layers (more concretely, a pooling sub-layer and a contrast normalization sub-layer) in each layer, as first proposed in the "neocognitron" network [53], and by training the circuit through unsupervised autoencoder learning with large amounts of image data (Figure 1c) [54]. The optimal stimulus for the cat-recognizing neuron, called "Google's cat," gained significant attention because the AI circuit autonomously acquired the concept of a cat without being provided any labeled information about cats during training. This result demonstrated the remarkable potential of the deep autoencoder to autonomously learn not only feature extraction of visual information but also object recognition and classification.

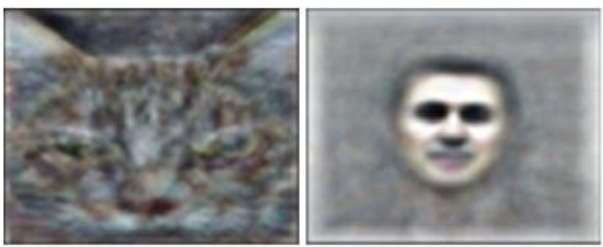

**Figure 1c.** The optimal stimuli that most strongly activated two representative neurons in the highest layer of the deep-layered (hierarchical) autoencoder. Left: The best activation stimulus for one neuron was this cat-like feature (Google's cat); this neuron is interpreted as a cat-classification neuron. Right: A human-face-classification neuron, whose best stimulus was a human facial feature, also emerged. Circuit structure: A deep circuit consisting of three repetitions of the receptive-field learning sub-layer (Figure 1a, red) combined with a pooling sub-layer and a contrast-normalization sub-layer. Input/output: A deep-layered structure where the signal compressed in a lower layer is used as the input for the next layer. Learning: The three receptive field sub-layers are trained sequentially, one layer at a time, starting from the lowest layer, to reconstruct the input signal (greedy layer-wise training). Panel **c** extracted from YouTube video of Peter Norvig's talk to the 2012 Singularity summit.

**Shift of image-recognition AI from deep-layered autoencoders to convolutional neural networks**

In 2012, a pivotal event occurred at a competition for image-recognition AI. While the mainstream methods for image recognition at the time (methods without neural networks) were unable to surpass the 75% correct rate barrier, AlexNet, created by the Hinton group using convolutional neural network (CNN) architecture, achieved an 84% correct rate and won by a wide margin (Supplementary Fig. 1). After that, the mainstream of object recognition AI shifted to CNNs, and AlexNet became the trigger for the subsequent widespread popularity of deep-layered neural networks, known as deep learning [55-57].

AlexNet and subsequent CNNs have a structure quite similar to deep autoencoders, with repeated layers of receptive-field processing and tolerance sub-layers, but there are two major differences. The first difference is that CNNs employ shared receptive-field processing across different positions in the image (see Figure 1d-f). This change significantly reduced the number of parameters to be trained and, for the first time, enabled easy training of deep layers. However, not sharing receptive-field processing is more biologically plausible and can yield greater invariance and robustness [54,58]. The second difference is that CNNs rely solely on supervised learning with correct labels (made by humans) and include no unsupervised learning. Since the goal of AI object recognition is to classify/label the input image (e.g., "This image is a dog."), deep autoencoders that undergo unsupervised pre-training require an additional output layer for labeling to be added to the last layer that is trained by supervised learning (fine-tuning), so that the circuit can generate the label output [59,60]. By contrast, CNNs do not require pre-training and complete their training with supervised learning alone. While this was seen as an advantage of CNNs, it diverges from the learning process in the visual system of the brain, which progresses almost entirely without correct labels [15,48].

The information processing in AlexNet and deep autoencoders contains similarities and differences (Figure 1d) [61-64]. In both, neurons in the lowest layer acquire receptive fields for line segments or local stripes, while neurons in the highest layer acquire specific responses to higher-level features (Figures 1b-c and e-f). However, a major difference is that in the highest convolution layer of AlexNet, neurons emerged that recognized cat faces, which were included in the labeled training data (i.e., in the instruction signals), whereas neurons that specifically recognized human faces—which were not included in the instruction signals—did not appear, unlike in the deep autoencoder [61,62]. This lack of generalization in AlexNet aligns with the observation that supervised learning circuits tend to develop information processing that is specialized to the instruction signals and ignore information not included in the instruction signals [17-20]. Furthermore, the emergence of human-face-detecting neurons in CNNs trained with unsupervised learning suggests that the difference in information processing is due to the difference in learning methods rather than differences in the circuit architecture (for intermediate stages of information processing, refer to Supplementary Fig. 2) [65].

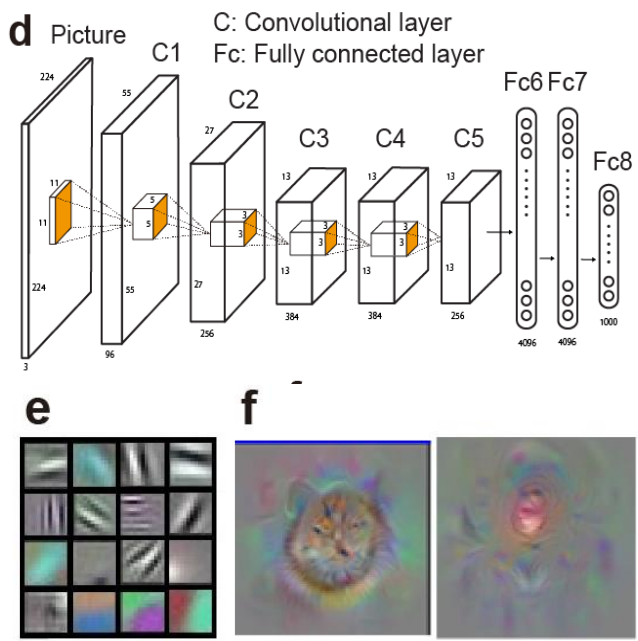

**Figure 1d-f.** The architecture and image processing of a representative CNN, AlexNet (the winning model in a 2012 object-recognition competition). In CNNs, local receptive fields are shared across various positions in an image. In the autoencoder, horizontal stripe-detecting neurons can be found at any location in the visual field (red boxes in Figure 1b): if the local information processing can be shifted in the visual field, these neurons can be regarded as applying nearly identical information processing to different local areas. Therefore, in CNNs, this local information processing is treated as identical, to share synaptic weights. For practical purposes, the neuronal processing of local horizontal stripe detection is applied via parallel sliding the receptive field to any position in the visual field. This sliding operation of the same local processing is termed convolution in mathematics and is the origin of the name "convolutional neural network". **d**, The circuit structure of AlexNet: 5 convolutional layers followed by 3 fully connected layers (i.e., classic neural-network layers). Input/output: The input is a color image (with 3 RGB channels). Corresponding to the image-classification task with 1,000 categories (e.g., "car"), the output is the probability that the input image belongs to each of the 1,000 categories, represented by the signal intensities of 1,000 neurons (Fc8). Learning: Supervised learning with human-made labels (from the ImageNet database). **e**, In the first layer of AlexNet, local-stripe-detection neurons similar to those in Figure 1b emerged. (from). **f**, In the final (5th) convolution layer in AlexNet, neurons that extract abstract features emerged. Neurons recognized a cat face (left) and features common to human/cat faces (right). Panel **d** adapted with permission from Kitazawa (2020) and Ramsundar & Zadeh (2018). Panels **e,f** adapted from Yosinski J. et al. (2015) (CC BY-NC-SA 3.0).

**Direct comparison of object-recognition processing by the brain and AI**

Inspired by advances in object-recognition AI, Yanis and colleagues conducted the first direct quantitative comparison between information processing in the brain and in AI [1,66-68]. They presented the same eight-category images to both a CNN trained with supervised learning and to monkeys, then compared the response signals in the AI and in visual neocortices. The information in higher visual neocortex was accurately reproduced by a linear combination of signals from the higher layers of the CNN. This suggests that the deep-layered structures of the CNN and visual neocortices produce a similar flow of information processing.

      Subsequently, a CNN trained in object recognition through unsupervised learning was compared with the visual neocortex, and similarities were observed between information processing in the lower layers of the CNN and lower visual neocortex, as well as in the higher layers of the CNN and higher visual neocortex [48,69]. Interestingly, when comparing the recognition accuracy of different learning methods, the highest accuracy was achieved by supervised learning and the second highest accuracy was achieved by contrastive learning, an unsupervised-learning method, whereas an

autoencoder produced considerably lower accuracy. In contrastive learning, neuronal activity for the current input image is compared with the neuronal activities for thousands of past input images and the synaptic weights are updated such that neural representations (i.e., neural response patterns) for the same object become more similar and neural representations for different objects become more different. This type of learning has gained popularity since 2020. To perform contrastive learning without human-made labels, several approaches have been proposed (For example, one method uses cluster analysis to regard the cluster which the neural response to the current image belongs to as the cluster of responses to the same object. Another method regards the responses to derivative images of the current image, such as inverted or grayscale images, as the responses to the same object). While the power of contrastive learning is impressive, comparing current activity with the responses to thousands of past images seems to lack physiologically feasibility, and there is no evidence yet that such learning is performed in the brain. On the other hand, the lower accuracy of the autoencoder was likely due to their use of a pre-2006 training method in which all layers are trained at once. We await future research on a direct comparison between neural activity in the visual neocortex and high-performing, biologically plausible AI circuits trained with unsupervised learning, such as unsupervised autoencoders. (The closest reported similarity to unsupervised learning AI is in auditory cortex and is discussed in section 1.2.)

**Evolution of circuit computation theory of the visual neocortex: Extension of autoencoders to prediction-error-learning RNNs**

As we have seen, deep autoencoder circuits trained through unsupervised learning have evolved from theories of brain circuits. When trained with natural images, the autoencoders develop Gabor-filter-like neurons in lower layers, and high-level feature-detection neurons in higher layers, just like in the brain. In this sense, autoencoders have succeeded in theoretically reproducing certain aspects of the circuit computations of the visual neocortex. However, deep autoencoders differ from the brain in several critical points. First, deep autoencoders are feedforward circuits, whereas the brain is characterized by abundant complex recurrent connections [70-73]. Second, because of this feedforward structure, autoencoders cannot integrate past and current information, making them incapable of processing dynamic or sequential visual images, such as videos, thereby differing from neocortical processing [14,71,74-84] (3D-CNN and vision transformer circuits will be discussed later in this section). To address this mismatch, adding recurrent connections to an autoencoder to form a recurrent neural network (RNN) circuit, which allows for the integration of information over time, would provide a more accurate theoretical replication of brain circuitry. For the input–output signals and learning of the RNN, we propose "prediction-error learning"—where the RNN predicts future inputs on the basis of past inputs and learns from the prediction errors—for the following reasons. First, prediction-error learning allows for numerous cycles of unsupervised learning, making it a promising learning method for the brain, which requires extensive unsupervised learning [1,20,85-87]. Second, prediction-error learning facilitates the separation of background and foreground (objects) and the learning of physical laws, which are difficult for autoencoders lacking temporal information [20,88-90]. Experiencing changes in the 2D projections of objects and their motion over time is a critical signal for learning about 3D object structure and physical laws [47,72,91-101]. We therefore propose that an "RNN circuit that predicts future input signals on the basis of past input signals and learns from prediction errors" (hereinafter, a "prediction-error-learning RNN") is a more biologically plausible theoretical circuit of the neocortex than a deep autoencoder.

    A circuit that fits the criteria of the "prediction-error-learning RNN" has already been proposed as a theoretical circuit for the neocortex: the predictive coding circuit (Figure 2a) [29,102-104]. Predictive coding was originally proposed to explain receptive field properties that could not be explained by classical theories of the visual neocortex. A growing body of experimental evidence supports this theory [8,103,105]. The theory has also been generalized to other sensory areas, including the somatosensory and auditory neocortices [8,106-110]. The predictive coding circuit resembles the autoencoder circuit but differs in three major points (compare Figures 1a and 2a). First, the autoencoder circuit receives input only once for the current visual image, whereas the predictive coding circuit receives input continuously in a sequential manner. Second, the predictive coding circuit includes a loop pathway with a delay at the compressed-prediction neuron, enabling temporal integration of past compressed predictions with the new input. Third, instead of directly transmitting input information to the compressed-prediction neurons, the predictive coding circuit calculates the difference between the prediction and the input and transmits only the prediction error (hence the name "predictive coding"). In the autoencoder, the error information is computed solely for learning (Figure 1a, green), whereas in the predictive coding circuit, the error is generated within the main circuit and directly contributes to the output. Fourth, while the autoencoder uses the current input as the correct information to reconstruct, the predictive coding circuit uses the future input as the correct information for prediction-error learning. When the predictive coding circuit processes a static image, the past and new inputs are identical, and the input information is compressed by the compressed-prediction neurons and reconstructed by the prediction neurons, which is equivalent to the autoencoder process. In other words, the predictive coding circuit can be interpreted as a time-oriented extension of the autoencoder to an RNN.

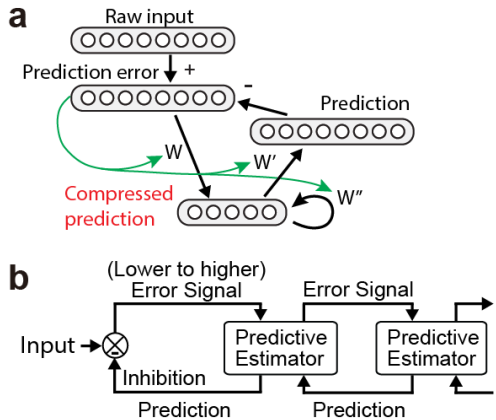

**Figure 2. Prediction-error-learning RNNs as a processing theory for dynamic visual information in the neocortex.**

**a, b**, Predictive coding circuit. **a**, The error signal (from the prediction-error neurons) is central not only for learning but also for information flow in the circuit. **a**, To establish a deep-layered structure, each layer (corresponding to the entire circuit in **a**) sends a copy of the error signal to the next higher layer and receives a copy of the compressed prediction from that higher layer. Circuit structure: The predictive coding circuit computes the difference between the input and the prediction. This prediction error signal is sent to and compressed by the compressed-prediction neurons, which have a recurrent connection and integrate past and current information. The prediction neurons restore the original data dimensions and predict the next input. Input/output: The input is a 2D visual image. The primary output is the prediction of the next input. With a deep-layered structure, secondary outputs are the prediction error signal that is sent to the next higher layer and the compressed prediction sent to the lower layer. Learning: The synaptic weights of the circuit are updated to minimize the prediction-error signal (the learning signal is shown in green). Panel **b** adapted from Rao & Ballard (1999) with permission from Springer Nature.

Furthermore, predictive coding circuits can be stacked to form deep-layered structure to incorporate the reciprocal connections between the lower and higher visual neocortices [12,103,104,111]. In a deep-layered predictive coding circuit, each layer sends the error signal of the prediction-error neurons to the next higher layer (Figure 2b) and receives the compressed prediction signal from that higher layer, which serves as an additional input signal to the compressed-prediction neurons (see also Figure 2c). In other words, a signal that was not predicted in the lower layer is transmitted to the higher layer to become the new prediction target (an unpredicted "newsworthy" signal) [29], and the higher-level prediction signal is fed back as a top-down signal to the lower layer. Moreover, the "free energy principle" theory, which abstractly and mathematically formalizes the prediction errors in predictive coding circuits as "entropy," suggests that the dynamics of the world/environment, which usually has deep-layered structure, can be modeled by mapping/embedding the dynamics onto a deep predictive coding circuit [12]. The deep predictive coding circuit generates prediction errors at each layer and prediction errors are consumed at each layer for learning, thus it can be viewed as a deep-layered prediction-error-learning RNN.

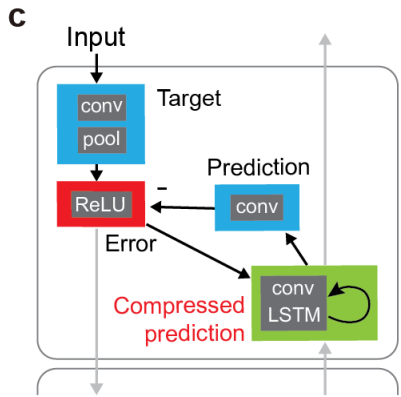

**Figure 2c.** Deep predictive coding network (PredNet). This network incorporates the core circuitry of the predictive coding circuit shown in Figure 2a with additional convolution sublayers, a pooling sublayer, and a long short-term memory (LSTM) unit. The LSTM integrates temporal information via a loop pathway. The inputs and outputs across layers are depicted with gray arrows. ReLU, rectified linear unit.

**Hints on how the brain achieves synaptic updates equivalent to backpropagation**

Circuit architectures that locally generate and locally consume error signals, such as deep autoencoders and predictive coding circuits, could be the key to solving the problem that it is not biologically plausible to assume backpropagation in the brain. The brain achieves hierarchical/deep-layered visual processing in ways that are comparable to AI, but it remains largely unknown how the updated synaptic weights in the brain are linked to improvements in the final output (i.e., how the brain accomplishes the credit-assignment problem). During supervised learning in deep-layered AI circuits such as CNNs, the error information generated in the final layer is necessary for synaptic updates in the first layer. Backpropagation solves this problem in a mathematically elegant way, but no equivalent signal for backpropagation has been confirmed to exist in the brain (i.e., the top–down signals from higher layers to lower layers in the brain match backpropagation only in the signal direction: the information conveyed is quite different) [4]. In deep autoencoders or predictive coding circuits, however, the error signal generated by each layer is consumed only within the same layer, so there is no need to carry the error signal far across layers. Transporting error signals over long distances can cause issues like the gradient exploding or vanishing, where the backpropagated signal can become too large or approach zero, often leading to learning failures. Assuming that deep autoencoders and the predictive coding circuits are plausible theoretical circuits of the neocortex, we can infer that the brain—which cannot afford to fail to learn—avoids such learning problems by generating error signals locally and consuming them locally.

**Prediction-error-learning RNNs among video processing AI: PredNet, CNN-RNN**

Given that prediction-error-learning RNN circuits are superior to autoencoders in processing dynamic signals (such as videos, as opposed to still images), we here focus on AI circuits designed for video analysis in our comparison of AI and the brain. Notably, an AI circuit named the Deep Predictive Coding Network (PredNet) was inspired by deep-layered predictive coding theory in the neocortex (Figure 2c) [70,72]. PredNet is a practical implementation of a deep predictive coding circuit, but it differs from neocortical circuitry in its use of LSTM for the RNN component because the detailed interconnection patterns between different cell types in the neocortex are still largely unknown. Although LSTM is an excellent RNN circuit commonly used in AI to achieve integration of temporally distant past and current information and efficient backpropagation, it is not designed to replicate the neocortical circuit structure or the attention mechanism (discussed later). On the other hand, in terms of input/output signals and the learning process, PredNet faithfully implements deep predictive coding and adheres to neocortical theory by predicting future inputs from past inputs and learning from the prediction errors. PredNet achieved high prediction accuracy for the next frame of rotating objects in a 3D environment (Figure 2d). Furthermore, PredNet developed internal representations that efficiently coded object parameters (e.g., identity, view, rotation speed) in the intermediate layer. This demonstrates that PredNet, aligned with the circuit theory of the visual neocortex, replicates certain functional aspects of the neocortex that autoencoders could not.

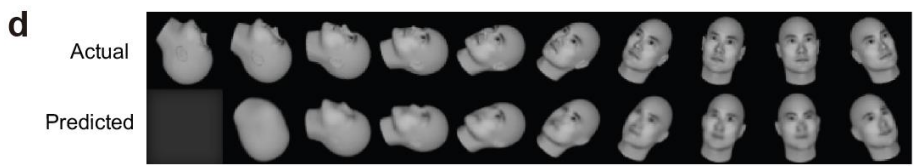

**Figure 2d**. Prediction by PredNet of the next frame in a sequence of 2D images (a face rotating in 3D space). PredNet successfully achieved higher prediction accuracy than either a CNN-RNN or an autoencoder tested by the authors. Panels **c,d** adapted with permission from Lotter W. et al. (2016).

In addition to PredNet, CNN-RNNs form another type of AI circuit within the "prediction-error-learning RNN" category. Here, we particularly highlight a CNN-RNN that has succeeded in acquiring models of video game worlds through the processing of game video information (Figure 2e,f) [112,113]. Traditional AI circuits for playing video games did not separate the video processing unit from the controller unit (i.e., the action selection unit to select the button to press), and the video processing unit was trained via backpropagation of the reinforcement learning signal from the controller unit. This approach produced slow learning and limited the trainable size of circuits (e.g., DQN, see 2.2). Therefore, instead, Ha and Schmidhuber trained a large-scale CNN-RNN as the video processing unit—independent of the controller unit—using unsupervised prediction-error learning, in which the CNN-RNN predicted the future frame from past frames (Schmidhuber is a proponent of LSTM). Interestingly, this CNN-RNN acquired "world models" of the games, including information on game events and event transitions. The combination of CNN and RNN was chosen because the CNN first compresses the visual information, making it efficient for the subsequent RNN to integrate information in the time direction. The world models acquired by this circuit proved so effective and powerful that the CNN-RNN was able to provide virtual games within the world models and thus improve the capabilities of the controller unit without any actual game inputs (below).

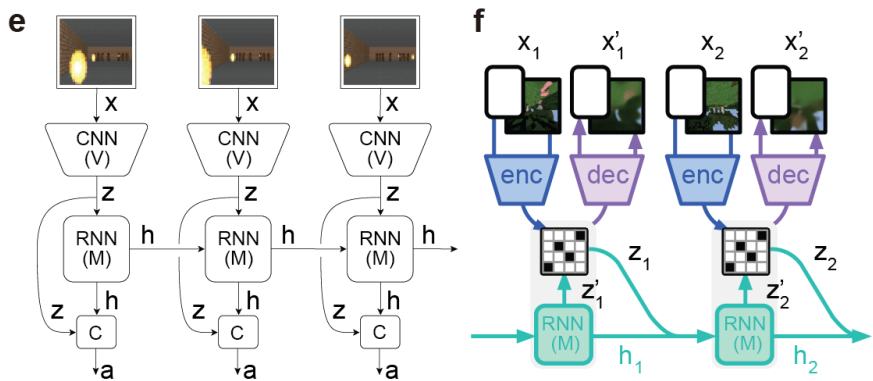

**Figure 2e,f**. Video-game processing-controller network for acquiring world models. **e,** This network consists of a CNN and an RNN, where the CNN is called the vision (V) module and the RNN is called the memory (M) module. V and M modules (CNN-RNN) are first trained through prediction-error learning to acquire the world models of games. After this, the controller (C) module learns to control playing the video game. The controller module is classified as a model-based reinforcement learning unit, discussed later. Note that the loop connection (within RNN unit) is depicted as unfolded in the time direction. **f,** Learning process of the CNN-RNN.

Circuit structure: The V and M modules are a CNN-RNN (LSTM) circuit. Input/output: The CNN, with an encoder–decoder structure, receives a video-game frame (x) as input and outputs the reconstructed frame (x'; in **f**). The secondary output of the CNN is the compressed information (z), which is fed to the RNN and game controller. The RNN takes the compressed frame information (z) from the CNN as input and outputs the prediction of the next input (z'). The recurrent information (h) in the RNN, its secondary output, is sent to the game controller (C module). Learning: The CNN functions as an autoencoder, learning to encode the input frame information into a compressed representation and then decode it to reconstruct the same frame at the output. (More precisely, variational autoencoder learning was implemented to learn similar/clustered compressed signals for images of similar objects/meanings.) The RNN learns through prediction-error learning, where it predicts the next compressed information (z') from the CNN from past compressed frames (z), updating its synaptic weights according to the prediction error. Panels **e,f** adapted from Ha & Schmidhuber (2018) (CC BY 4.0). and Hafner et al. (2023) (CC BY 4.0).

Although CNN-RNN circuits excel in online sequential processing of video frames, practically speaking, there is greater demand for AI offline analysis of recorded videos (such as YouTube videos) than for online processing. In offline video analysis, RNNs have been avoided because of the computational cost (i.e., the difficulty of efficient parallel computation with GPUs), and the popular lineage is 3D CNNs (popular from 2014–2020), which process 3D input with 2D pixel information plus 1D time information, and the vision transformer (ViT) series (popular from 2020 to the present), which replace CNNs with transformer circuits, while retaining similar input dimensions. These circuits process all frames at once by feedforward-circuit computation, and the typical learning method is supervised learning with correct human-made labels (e.g., "a moving car"). Therefore, the circuit structure, input–output, and learning of 3D-CNN and vision transformer circuits have few similarities with the brain. However, combining prediction-error learning, where parts of video frames or image pixels are masked, with supervised learning has been shown to improve their performance; this demonstrates the power of prediction-error learning [114].

Finally, we introduce an AI circuit that achieved neural representation of 3D information—an ultimate goal of visual processing—using only unsupervised prediction-error learning (Figure 2g,h) [115]. The AI circuit received inputs comprising 2D images of multiple objects taken from various viewpoints in 3D space, and it was trained through prediction-error learning to predict the 2D appearance of these 3D objects from a new, unexperienced viewpoint. The circuit achieved high prediction accuracy of 2D images from unexperienced viewpoints, strongly suggesting that the internal signal of the circuit acquired neural representations of all the information necessary for prediction, (such as the color and shape of objects and the positional relationships between objects)—in other words, a neural representation of the 3D environment. While the architecture of this AI circuit is a type of CNN-RNN, it incorporates many elements representing Gaussian distributions, which differentiates it from brain circuits. Since this is not the only circuit capable of achieving representation of 3D information [115], it will be worthwhile to create a more biologically plausible artificial circuit to reproduce the acquisition of 3D neural representations in the brain.

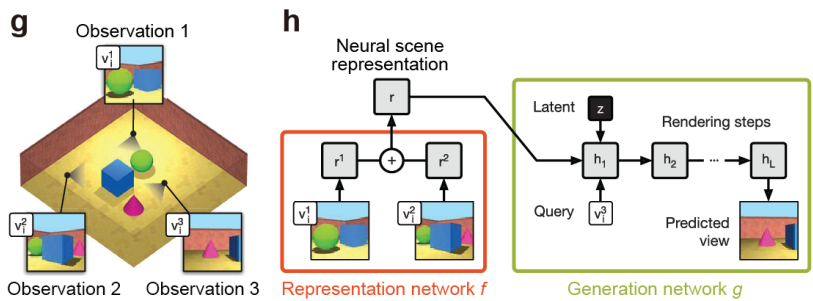

**Figure 2g,h**. A CNN-RNN that acquired an internal representation of a 3D environment and is capable of predicting 2D images from unexperienced viewpoints. **g**, 2D images corresponding to different viewpoints within a 3D environment. **h**, Circuit diagram illustrating the process of receiving observations 1 and 2 as inputs to predict observation 3. The representation network (red enclosure) is a CNN that compresses the input images, while the generation network (green) is an RNN (skip-convolutional LSTM) that generates a 2D image to predict the view. Learning is driven by the predictive error, and the entire network is trained through backpropagation of the error. Panels **g,h** adapted from Eslami SMA et al. (2018) with permission from AAAS.

PredNet and these CNN-RNNs exemplify that a prediction-error-learning RNN circuit can acquire world models that represent latent information, such as game event transitions and 3D environments, by abstracting the necessary information for prediction. The brain also acquires internal models of the external world through experience and utilizes the models for information processing across various domains in the neocortex [12,24-35] and the cerebellum [14,36-43]. This similarity between AI and the brain is very intriguing, as it suggests that by creating a prediction-error-learning RNN circuit replicating the brain and examining how it learns and processes information, we could significantly advance our understanding of how the brain acquires world models and represents latent information, mechanisms that remain largely unknown.

## 1.2 Cognitive Processing in Sensory Systems

Sensory information processing includes highly sophisticated cognitive processing, such as human-characteristic language processing. Since voice recognition is commonly found in other animals rather than being unique to humans, our particular focus here is on sensory language processing of sentence comprehension after word recognition.

**A theory of circuit computation in the cerebellum for language processing: A three-layer RNN circuit with prediction-error learning for next-word prediction**

While the cerebellum is often considered a center for motor control, it is also responsible for a wide range of cognitive functions, including language processing [116-123]. The right lateral cerebellum (Crus I/II) is involved in two important non-motor language-processing functions: predicting the next word in a sentence [124-127] and grammatical processing, particularly syntactic processing [121,128-131].

In cerebellar language processing, the three elements of circuit computation (circuit structure, input/output, and learning) are considered as follows: Regarding the circuit structure, the details of the uniform cytoarchitecture of the cerebellum are well studied [2,123,132-134]. Although the cerebellum is often described as a typical feedforward circuit, recent studies have revealed that there are abundant feedback projections from the cerebellar nucleus neurons (the output neurons; Figure 3a) to the granule cells (the input neurons), including both direct projections [133,135-137] and indirect projections through the pontine or brain stem [138-141]. These feedback projections are essential for the predictive functions of the cerebellum [137,142,143]. Regarding the input and output signals of the circuit, historical cerebellar research has revealed that the function of the cerebellum is to create internal brain models of the world and predict the future (the widely-accepted "cerebellar internal model" theory discussed below) [14,36-43]. Extending this to language processing, it has been proposed that next-word prediction is achieved by the cerebellum receiving sequential word inputs and generating a prediction (output) of the next word [124,125,144]. Regarding learning, it has been proposed that the inferior olive, which sends signals essential for cerebellar learning, compares the next-word prediction (the output of the cerebellum) with the actual next word to calculate the prediction error, which is then used to update synaptic weights to improve future predictions [122,144].

Using these elements as the basis, and with the aim of reproducing cerebellar language functions, we created an artificial neural network that imitates the cerebellar circuit structure, including the connection patterns among different cell types in the cerebellum. We connected the three layers of input, Purkinje, and output neurons with feedforward and feedback pathways according to recent knowledge (outlined above) to create a three-layered RNN circuit (Figure 3a). Interestingly, when the cerebellar ANN was trained to perform next-word prediction, not only did the output of the circuit acquire the ability to predict the next word in a sentence (Figure 3b, red arrow), but the intermediate layer of the word-prediction circuit (Purkinje cells) spontaneously acquired another cerebellar language function, syntactic processing (Figure 3b, blue) [145]. To the best of our knowledge, this is the first brain-imitating artificial circuit that aligns with neuroscientific findings for all three of circuit structure, input/output, and learning and that reproduces sophisticated human-characteristic cognitive functions. This result demonstrates that the two cerebellar language functions (next-word prediction and syntactic processing) can be viewed as a unified computation of a single circuit. Given the uniform cytoarchitecture of cerebellum, this result also indicates the possibility of unified comprehension of two distinctive cerebellar functions of predicting future events (the function of the cerebellar internal model) and extracting temporal features from sequential events (sequence processing). In other words, by viewing the cerebellum as a three-layer RNN circuit that predicts future words from past words and learns via prediction errors, we provided a unified explanation of the two cerebellar language functions as well as insights into the mechanisms underlying universal cerebellar circuit computation.

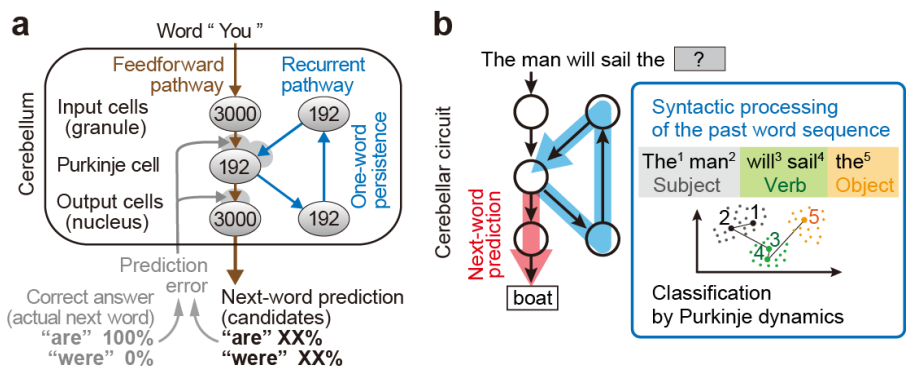

**Figure 3. Reproduction of cerebellar language functions by a three-layer artificial RNN imitating the cerebellum.**

**a**, A circuit was trained in a cerebellar language function—predicting the next word from a sequence of past words—through prediction-error learning. Each of the 3000 input neurons (granule cells) represents one word in the 3000-word set (sparse coding). Purkinje cells integrate the current word information in the feedforward pathway (brown) with the previous word-sequence information in the recurrent pathway (blue). The firing rates of the output neurons (cerebellar nucleus neurons) represent the probabilities that each of the 3000 words is the next word, forming the next-word prediction. **b**, When the output neurons were successfully trained to predict the next word (red arrow), a circuit for syntactic processing (specifically, classification of words as subject, verb, or object, another language function of the cerebellum) spontaneously emerged upstream of the output cells (blue arrow).

Circuit structure: A three-layer RNN circuit that replicates the physiological connectivity of the three cell types of the cerebellum. The cerebellum receives the prediction-error signal from the inferior olive through the climbing fiber pathway; this signal pathway is essential for cerebellar learning. This dedicated learning-signal pathway is a feature of cerebellum, distinct from the neocortex. Input/output: The circuit receives word information sequentially and the primary output is a prediction of the next word; syntactic information is the secondary output. Learning: The circuit performs prediction-error learning for next-word prediction. Figure adapted from Ohmae K. and Ohmae S. (2024) (CC BY 4.0).

While prediction-error learning in the cerebellum has traditionally been called "supervised learning" in the neuroscience field, here we follow recent terminology and refer to prediction-error learning in general (including that of the cerebellum) as "unsupervised learning." (Supervised learning refers to learning in which the correct signal explicitly includes the information that the circuit should learn.)

**Circuit computation in the neocortex for language processing: A deep-layered RNN circuit that predicts the next word and learns through the prediction error**

Compared with the cerebellum, neocortical circuits contain a wider variety of cell types and have complex multiple loops connecting these cell types [14,74-78,80,81,83,84]. Given that the same circuit architecture in the neocortex is used for various types of information processing [11,14], the neocortical language circuitry is similar in structure to the circuitry in the visual neocortices. Like visual processing, language processing occurs hierarchically in the neocortex [146-152] and can be summarized as hierarchical processing by deep-layered RNNs with complex recurrent connections. With regard to learning and how the brain acquires language, the innate theory of language (in which basic language functions are innate and genetic) has historically been favored [153-155], but recent evidence has challenged this theory, including simulation experiments demonstrating that language functions that had been assumed impossible to learn can in fact be acquired through biological postnatal experiences and unsupervised learning [156,157]. In this way, although room for controversy remains over the extent of innate language knowledge in the brain prior to learning, there is abundant evidence of similarities in information processing between the neocortex after language acquisition and AI trained on word prediction through prediction-error learning. In 2021, a functional MRI study reported that among recent high-performance AIs, the one with signals most similar to those of the neocortical language area was a prediction-error-learning AI [158]. In 2022, a human electrocorticography study demonstrated that that the signals in the broad neocortical language area contain both word-prediction and prediction-error information [159]. Additionally, electroencephalographic (EEG) and magnetoencephalographic (MEG) recordings indicated that language comprehension is supported by hierarchical predictions like linguistic AI [160]. In 2023, another human electrocorticography study found that the signals in high-order auditory neocortex (in the superior temporal gyrus) more closely resemble the signals of prediction-error-learning AI than those of supervised-learning AI [150]. In 2024, human single-neuron recordings revealed that word prediction contributes to word representation by neurons in the prefrontal cortex [161]. Taken together, these findings strongly support the view that, after language acquisition, language processing in the neocortex is conducted by a deep-layered RNN circuit that predicts future words from past words and learns via prediction errors.

**The lineage of language processing AI: deep-layered RNN (GNMT), transformer circuits (BERT, GPT)**

In the field of language-processing AI, artificial RNNs have been highly anticipated since the early 2010s [156,162,163]. A notable example is Google Neural Machine Translation (GNMT), which was adopted by Google Translate in 2016 and dramatically improved translation accuracy over previous methods [164,165]. GNMT employs a deep-layered RNN circuit that uses LSTM (Figure 4a) and consists of an encoder–decoder. However, while the deep-layered predictive coding circuit (Figure 2a-c) has an encoder–decoder in each layer, GNMT has an eight-layer encoder unit followed by an eight-layer decoder unit, so the encoder–decoder combination exists only once. As a result, whereas the predictive coding circuit calculates the prediction error at each layer, GNMT makes word predictions and calculates the prediction error only in the final, 16th, layer and propagates this error signal to all layers by backpropagation. Furthermore, GNMT relies on supervised learning, where human-made translation examples are used as the correct targets. Thus, while GNMT is similar to the neocortex because it is a deep-layered RNN, it differs in that prediction errors are generated only in the final layer and propagated to all layers by backpropagation, and in that it is trained through supervised learning. Nevertheless, the GNMT encoder can convert a sentence in one language into a semantic vector (the population activity of a group of neurons), and the decoder, upon receiving this vector, can generate a sentence with the same meaning in another language with near-human precision. This demonstrates that the GNMT encoder, an artificial deep-layered RNN, has the capacity to convert sentences into semantic population-coding vectors at a level close to that of humans, which was groundbreaking from the neuroscience perspective (also see seq2seq and seq2vec) [166,167].

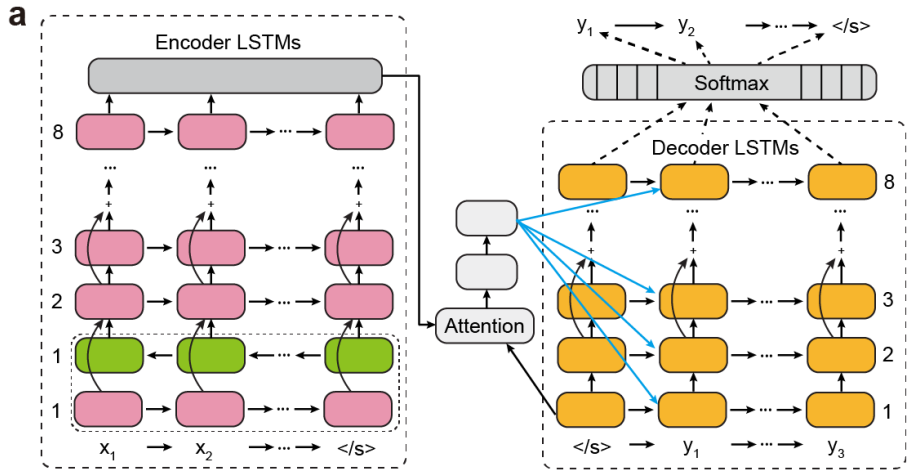

**Figure 4. Language processing AI and analogy to the neocortex.**

**a,** The translation circuit of Google Neural Machine Translation (GNMT). This circuit consists of encoder, attention, and decoder units. Left, The encoder is a stacked-LSTM circuit that sequentially receives the words of a source sentence and integrates them to generate the signal that represents the meaning of the entire sentence. The LSTMs are bidirectional, containing units that integrate information sequentially from the beginning of the sentence to the end (in red) and from the end back to the beginning (in green). As a result, the representation for even the first word can include the context of the sentence. The encoder has eight layers (one bidirectional layer plus seven unidirectional layers), and the outputs of the final layer are combined to generate a vector representing the meaning of the entire sentence. Center: The attention unit is a two-layer neural network that receives the most recently translated word from the decoder (arrow from the decoder; only the signal flow during the second word [y2] translation is illustrated) and the vector representation of the source sentence from the encoder (arrow from the encoder), and then creates a weight-coefficient signal indicating which word in the source sentence should be focused on (i.e., an attention signal; made by an attention mechanism called the additive attention; discussed later). The attention unit then provides this weight-coefficient multiplied source-sentence signal to each layer of the decoder (blue arrows to the decoder). The decoder is an eight-layer LSTM that generates the probabilities of each word being the next translated word sequentially (converted to probabilities by Softmax).

Circuit structure: The circuit consists of an eight-layer LSTM encoder, an attention unit, and an eight-layer LSTM decoder. Input/output: The input is a sentence in the source language, fed into the LSTM word by word, and the output is the translated sentence in the target language, created by sequentially generating the next word. Learning: The circuit is trained through supervised learning with human-made translation examples as the correct targets. The error signal (calculated by comparing the generated word with the correct word) is delivered to the entire encoder–decoder system via backpropagation to update the synaptic weights. Panel **a** adapted from Ramsundar & Zadeh (2018) with permission from O'Reilly Media.

After GNMT, because of issues related to computational cost and speed (low efficiency of parallel processing with GPUs), AI language processing shifted away from the method of feeding words one by one into an RNN. Instead, the community moved toward the method of feeding an entire sentence into a feedforward circuit called a transformer [168,169]. This process differs from that of the brain, which receives sequential word-by-word input. But, interestingly, the prevailing learning method subsequently also shifted from supervised learning to unsupervised prediction-error learning, similar to learning in the brain.

The first to demonstrate the power and versatility of unsupervised prediction-error learning in a transformer circuit was Google's BERT (bidirectional encoder representations from transformers). BERT achieved top-level or best performance in 11 language tasks [18]. The pre-BERT language AIs (GNMT and its successors) suffered from the problem that supervised learning is not versatile (i.e., accuracy dropped significantly outside of the trained tasks), and the cost of preparing large training datasets is substantial. BERT solved this by first conducting large-scale unsupervised learning with an encoder–decoder circuit, then discarding the decoder unit and attaching a small output layer to the remaining encoder unit (which was able to convert sentences into semantic vectors, like GNMT), and performing additional small-scale supervised learning on the output layer for new language tasks. Thus, BERT needs different output layers for different tasks. With this method, BERT achieved strong performance across various language tasks. The circuit structure of BERT is a deep-layered transformer circuit, and the main learning method is unsupervised prediction-error learning that predicts a masked word. This prediction is different from next-word prediction in that it can also utilize words that come after the masked word.

The power of unsupervised prediction-error learning was demonstrated even more sensationally by the GPT (generative pre-trained transformer) series, including ChatGPT (Figure 4d) [170-172]. Especially since GPT-2, GPTs have been thoroughly trained in next-word prediction. When generating sentences, GPTs directly use the ability to sequentially select the word that follows the previous word sequence, learned by next-word prediction. This sentence-generation ability has allowed GPTs to solve various language tasks. (For example, in the task of scoring a product comment, when GPT is asked, "On a scale of 1 to 10, what is the positivity of this comment?", GPT can score it by sequentially selecting the most appropriate words to follow, like "This is an 8".)  The change in output method made GPTs versatile enough to solve various language tasks without any supervised learning, which had been necessary to train the output layers of pre-BERT AIs (e.g., the product-comment scoring task previously required a one-dimensional output layer for generating real numbers from 0 to 10). GPTs exhibited high performance across various language tasks, demonstrating that a transformer trained with unsupervised prediction-error learning is capable of acquiring remarkable language-comprehension abilities (the text-generation ability made famous by ChatGPT is discussed later in section 2.1). In recent years, AI has made great strides in supervised learning with CNNs and has moved away from brain learning. Interestingly, however, the latest AI learning is converging on prediction-error learning, which is similar to learning in the brain, and further development of AI by prediction-error learning is anticipated by the godfathers of deep learning [86,87].

The neocortex is a general-purpose language-processing circuit highly capable of solving various language tasks, and is anticipated to have similarities to GPT. In fact, a study examining which language-AI signals were closest to the signals in the neocortex found that recent language-AIs that used transformer circuits had higher similarity than one-generation-old AIs that used deep-layered RNNs (i.e., the successors to GNMT) [158]. Furthermore, in a comparison of transformer circuits with unsupervised learning, the signals from GPT trained by next-word prediction were closer to neocortical signals than those of BERT trained by masked-word prediction. This result suggests that, among all existing language AIs, GPTs in particular—transformer circuits trained with next-word prediction error—process language with some form of commonality to the neocortex [158].

**Similarities in processing between the neocortex and the transformer circuit**

It is notable that although transformers do not imitate the circuit structure of the neocortex, there is a high degree of similarity in the signals and mechanisms of information processing. There are also quantitative and qualitative similarities in the language-processing outputs: GPTs possess sophisticated language-processing capabilities comparable to those of the neocortex. In particular, as is widely known with ChatGPT, GPTs can instantly perform nearly all language tasks, such as translation, information provision, summarization, and scoring, based on natural language instruction (called "prompts") without any additional learning process (see below). In this sense, GPTs have a human-like adaptability to new tasks, setting them apart from conventional AI. As a result, even within highly varied interaction contexts, like those seen with ChatGPT, the text generated by GPTs is almost indistinguishable from human-written text [173]. In light of this, here we discuss the similarities in information processing between the neocortex and transformer circuits.

The first similarity is their high versatility. The neocortex processes information using circuits with nearly identical structures across sensory, motor, and cognitive functional domains [11,14]. Transformer circuits have excelled in various fields, achieving top accuracy in language processing (e.g., GPT, BERT, PaLM) and visual information processing (e.g., vision transformer), as well as excellent performance in video generation (e.g., DALL-E), suggesting their potential as a general-purpose circuit [20,168,174]. Second, through large-scale unsupervised prediction-error learning, powerful models of the external world, including visual world models and language models, can be built by both the neocortex [12,24-35] and the transformer [169-171,175,176]. Third, they both possess attentional processing mechanisms. The attention mechanism in the neocortex (specifically, top–down attention to spotlight specific information depending on context; for example, amplifying the signal intensity for certain words in an input sentence) is observed in a wide range of neocortical regions, such as the prefrontal and parietal areas. The transformer circuit was designed as an artificial circuit that realizes the top–down attention mechanism in the brain [16,177].

The differences lie in the circuit structure. To list these, first, transformers are feedforward circuits that receive inputs over time all at once (e.g., they receive an entire sequence of words simultaneously), whereas the neocortex operates as an RNN circuit, receiving inputs sequentially. However, this difference may not be fundamental, because any RNN can be represented by a time-unfolded feedforward circuit (e.g., Figure 2e,f), and the input to a transformer can be considered as input to an unfolded feedforward circuit [6]. Second, transformers lack a unit corresponding to an artificial neuron. Yet, this may also not be a major difference, as the transformer's attention mechanism (dot-product attention) can be reformulated as a two-layer neural network implementing additive attention (see the caption for Figure 4). Furthermore, the integration of information by attention mechanisms in transformers resembles that of CNNs, with the key difference being that while CNNs perform fixed-pattern integration of local inputs, transformers extend this to dynamic-pattern integration of distant inputs [178]. The similarities between visual neocortex and CNNs are increasingly recognized, and neither CNNs nor transformers reproduce the pattern of cell-to-cell connections in the neocortex; thus, it is not a major logical jump to replace CNNs with transformers as the comparison target.

Next, we elaborate on the processing mechanisms of transformer circuits in some detail. Transformer circuits have a very simple structure, and at their center is the attention mechanism designed in reference to the brain. This attention mechanism receives a copy of the signal currently being processed (the first input in Figure 4b) and generates an attention signal (Figure 4b, red arrow) that represents the degree of attention (coefficient) for each element of the target signal (the second input in Figure 4b). These attention coefficients are then multiplied with the respective elements of the target signal (Figure 4b, copy of the second input) to output a weighted sum in which the attended part is amplified. The attention signal in transformer circuits is calculated using only linear transformations and a dot product (Figure 4b, boxed section). Another important feature of the transformer attention mechanism is that, in addition to the standard attention described above, it can also generate self-attention. For example, in order to understand a specific word in a sentence, self-attention determines which part of the same sentence to attend to (e.g., to understand "it" in "Mary had a little lamb, and it was very white", we need to attend to "a little lamb"). This mechanism is called self-attention because the context for attention and the target of attention are contained within the same sentence (Figure 4c, the first and the second inputs). By stacking the self-attention layers and training this deep-layered circuit with prediction-error learning, GPT and the vision transformer achieved processing performance comparable to or better than human language and visual neocortices (Figure 4d). Thus, although the circuit structure of transformers is not biological, by imitating the neocortical attention mechanism of amplifying the signal of focus, transformers effectively implement both standard attention and self-attention, accomplishing general-purpose processing that is equal to or better than that of humans.

The questions of what kind of neocortical circuit computations underlie standard attention, whether the neocortex operates self-attention in addition to standard attention, and, if so, whether the standard attention mechanism extend to self-attention, remain very intriguing but unresolved questions. In the visual system of non-human primates, where the attention mechanism of the neocortex has been most studied, it is known that spatial attention amplifies the signal in the attended part of the visual field by increasing the gain of signal processing in that region. In terms of the output signal, the transformer attention mechanism reproduces this well.

However, the details of the neocortical circuit computation that realizes the attention mechanism remain to be elucidated. For example, there is controversy about whether attention-based signal selection occurs during early or late stages of sensory processing [179,180]. As a potential resolution to this controversy, one theory proposes that the stage at which signal selection occurs can be adjusted depending on the task demands [181]. Furthermore, the oculomotor network, which directs the gaze to a particular direction, is also involved in directing attention to that same direction (attention network theory)[182-185]. On the basis of these phenomena, numerous theories of the neocortical attention mechanism have been proposed [186-188]. The controversy between early and late selection theories implicitly assumes that there is only one attention mechanism in the brain; if this is the case, we need to expect a highly elusive mechanism that appears early or later in sensory processing and that shares processing with the oculomotor network. However, the functional similarities between the deep-layered transformer circuit and the neocortex suggest that the attention mechanism may in fact exist at each stage of information processing. Indeed, a human electrocorticography study demonstrated that the signals in high-order auditory cortex (in the superior temporal gyrus) correlate with the self-attention signals of a transformer circuit trained through prediction-error learning, directly suggesting the possibility of transformer-like attentional processing within neocortical circuitry [150]. In order to clarify the circuit computations underlying the diverse functions of the neocortex, it is necessary to investigate circuit-computation-level questions, such as whether attention mechanisms like those in transformer circuits are incorporated in the local circuit of the neocortex.

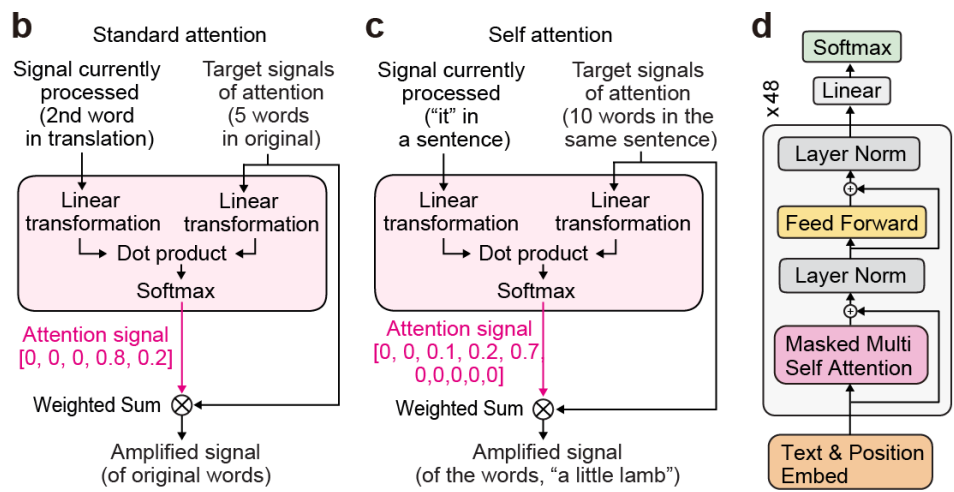

**Figure 4b-d.** Transformer attention circuit. **b**, Standard attention circuit. For simplicity, this will be explained with a concrete example. Consider the case where, during the translation of a sentence with five words, the signal for the second word that was most recently translated is used when generating the next word (input section). When selecting the next word, assuming that focusing on the corresponding word in the original sentence will lead to a more accurate and faithful translation, the attention mechanism (red box) generates an attention signal (degree of focus) for each word in the original sentence, with the sum being equal to 1 (red arrow; e.g., [0, 0, 0, 0.8, 0.2]). Next, this attention signal is used to weight the signal for each of the five words in the original sentence (with the attention signal serving as the weight coefficient), and the circuit outputs the original word information with the words of interest highlighted (output section). The first input is a high dimensional signal (e.g., a 512-dimensional signal), which provides the context of the current processing. The second input is the five 512-dimensional signals corresponding to the five original words. In the attention mechanism of transformer predecessors (e.g., GNMT), this input/output was learned in a two-layer neural network (called additive attention; see Figure 4a). By contrast, the dot-product attention mechanism in the transformer assumes that this input/output is, in this case, converting the second translated word (the first input) into candidates for the third word, transforming dimensions from the original language to the target language, and then calculating importance by comparing it with each of the five original words (the second input). It also assumes that the conversion to the next word and the dimensional transformation can be done only by linear transformation (rotation and scaling operations in 512 dimensions), and the importance calculation is done via a dot-product operation (similarity calculation) (inside the red box). This dot-product attention mechanism, owing to the rather aggressive assumptions, is much faster in computation and learning than additive

attention, yet, surprisingly, maintains comparable accuracy (to be precise, the linear transformation is applied to both input signals; in practice, to achieve higher performance, a multi-head attention process is employed, where eight linear transformations are performed in parallel to obtain eight parallel candidates of the attention signal). **c**, Self-attention mechanism. To understand "it" in a sentence, "Mary had a little lamb, and it was very white", the self-attention mechanism generates an attention signal that indicates which words in the sentence to attend to. The first and second inputs are signals from the same sentence (e.g., the signal for the word, "it", and the signal for the ten words of the sentence). The processing steps are the same as in **b**. If the word "it" corresponds to "a little lamb," the self-attention mechanism focuses on these three words and creates a signal to represent "a little lamb" by multiplying the attention coefficients (red arrow) and the word signals of the sentence (i.e., the copy of the second input) so that the output represents the detailed meaning of "it" (the information regarding "it" itself is also conveyed downstream by bypassing the attention mechanism; see the bypass path in **d**). **d**, Transformer circuit for GPT-2. Since GPT is a circuit that predicts the next word, it does not have an encoder–decoder structure like GNMT but has an integrated structure as shown in Figure 3a. Circuit structure: Repeating blocks mainly consisting of an attention unit and a two-layer feedforward neural network, with these blocks stacked 12 times in GPT-1 and 48 times in GPT-2. Input/output: The input is a sparse representation of about 50,000 words (as in Figure 3a), and the circuit input can receive text of up to 2048 words (including multiple sentences). The output is a probability representation of the next word candidate (as in Figure 3a). Learning: Unsupervised learning through next-word prediction. Panel **d** adopted from Wu Y. et.al (2024) (CC BY 4.0), with elements from Radford A. et al. (2018).

## 2. Motor Information Processing

Compared with sensory processing, less is known about what kinds of circuit structure, input/output, and learning are involved in motor processing. Even a relatively simple action such as brightening a room at dusk consists of several sequential stages: moving the body, reaching out and pressing a button on a lamp, and evaluating the resulting change in room brightness. More formally, cognitive theory suggests that actions consist of the following steps: To achieve the goal/intention, a sequence of actions is planned to fulfill a goal/intention (action planning), the action sequence is converted into a series of motor commands and executed (action execution), and the results of the actions are recognized and evaluated to refine future actions (action evaluation) (see the "seven stages of action" by Donald Norman for more details). This subdivision of action in cognitive science is also applied in robot control. In this section of our review, we discuss (1) planning of the action sequence (procedure), (2) evaluation and improvement of the action plan, and (3) generation of motor commands for individual actions and their evaluation and improvement. Finally, we highlight significant differences between robot control in AI and motor control in the brain.

## 2.1 Planning the Action Sequence (Cognitive Processing in Motor Systems)

**Sentence generation in motor language processing**

Continuing from the previous section on sensory language processing (particularly language comprehension), this section focuses on motor language processing, particularly language planning for sentence generation. At this stage, according to the speaker's intention and the goal of communication, appropriate words and grammatical structures are selected to form a sequence of words, generating a sentence (vocalization of the word sequence is not addressed here, as this is the execution stage of the action). Recent research has revealed that in the human brain, language planning is closely related to sensory language processing and language comprehension. Traditionally, it was believed that the sensory language area located near auditory neocortex (Wernicke's area) is involved in language comprehension, whereas the motor language area located near motor neocortex (Broca's area) is responsible for language planning. However, both sensory and motor language processing are now thought to be performed in significantly overlapping regions in and around Broca's area in the neocortex and in the right lateral cerebellum, which has strong reciprocal connections to Broca's area [124,125,129,147,151,152,189,190].

In GNMT, sensory language processing and motor language processing are carried out in separate circuits, which aligns with the traditional understanding of language in the brain. Concretely, the encoder unit of GNMT functions as the language-comprehension circuit that converts the meaning of a sentence into a high-dimensional vector (represented by an activity pattern of neurons), while the decoder unit is a language-planning circuit that generates a sentence in another language from the sentence vector (Figure 4a). On the other hand, in GPT, which is a transformer circuit trained through prediction-error learning, the circuits for sensory and motor language processing significantly overlap. GPT also has an encoder–decoder structure, but these are integrated without a clear boundary (Figure 4d). In terms of sensory language processing, GPT receives an input sentence and outputs a prediction of the next word. Through prediction-error learning, GPT acquires a model of the language world and the ability to comprehend sentences. Then, in terms of motor language processing, upon receiving a contextual input, the same circuit of GPT generates sentences. In this sense, GPT repurposes the word-selection mechanism of predictive processing straight into word selection for generation of sentences. In other words, GPT reopposes the word-prediction circuit to word generation. (This adaptation of prediction AI to information-processing and generative AI is common in other fields than language AI.) This overlap between sensory and motor processing in GPT aligns with our more recent understanding of the brain. This begs the question, does the brain similarly repurpose its word-prediction circuit to both sentence comprehension and sentence generation? For the cerebellum, the answer is likely to be "yes." The right lateral cerebellum is involved in both next-word prediction and syntactic processing, and a simulation study recently demonstrated that, through next-word prediction, the prediction circuit of the cerebellum can acquire a language model that enables syntactic processing [145]. Since the right lateral cerebellum is also involved in planning for sentence generation [121,191], it is likely that the cerebellar word-prediction circuit is also used for sentence generation. In the neocortex, high-order frontal regions involved in word prediction during listening also overlap with areas responsible for planning of sentence generation [158]. Furthermore, it has been proposed that the speech-perception system is critically involved in speech production (via the mirror neuron system, discussed later) [192-195]. This idea is further supported by human single-neuron recordings indicating that neurons involved in both listening and speaking exist in the prefrontal cortex [196]. Therefore, it is highly likely that prediction circuits in the neocortex are repurposed to sentence generation. The time has come to directly investigate whether the neocortex and cerebellum, like GPTs, utilize the word-selection mechanism acquired through word prediction to sentence-generation processing.

**Immediate adaptability to novel tasks in the motor language system**

The ability to, under appropriate instructions, learn and adapt quickly to a diverse array of entirely new tasks is a remarkable human characteristic rarely observed in other animals [197,198]. In the context of language generation, for instance, once given appropriate instructions, humans can immediately carry out a new task, whether it is translation, information provision, or product evaluation. Recent GPT language AIs (since GPT-3) have also achieved similar adaptive response capability, yielding the famous commercially available AI, ChatGPT (released in 2022) [169,173,199]. Concretely, by receiving instructions called prompts, GPTs can perform a wide variety of new tasks. As long as the tasks can be addressed through language-based responses, GPTs can respond with human-level quality (sometimes surpassing humans), even for quite unreasonable demands. This adaptability sets GPTs apart from traditional language AIs, which specialize in a single task (e.g., translation).

The mechanism that enables GPTs to adapt immediately to new tasks is considered to be as follows. The parameters of the GPT transformer circuit (corresponding to synaptic weights in neurons) are updated only during training and remain unchanged even when adapting to new tasks. However, when a prompt is given, the contextual information maintained in the self-attention mechanism of the transformer continuously sends a signal to change the input–output transformation of the entire circuit as if the parameters have been updated in response to the prompt [200,201]. This phenomenon, where the input–output transformation changes in response to the context of a given prompt, is called "in-context learning" [173,201,202]. The acquisition of in-context learning by GPT-3 surprised even its developers, as this capability was not explicitly requested in the circuit design or learning process. Since the main differences between GPT-3 and GPT-2 are the circuit size and the volume of training data, the emergence of in-context learning should be attributed to the scaling up of the language model [169,173].

Although the detailed mechanisms underlying the human brain's adaptability to novel tasks are still largely unknown, a mechanism has been proposed in which multiple simple processing circuits are prepared, a new task is broken down into multiple steps, the most suitable simple processing circuit is selected for each step, and the task is performed by switching between these circuits dynamically (the mixture-of-experts and MOSAIC mechanisms)[111,203]. This mechanism of dynamically switching processing circuits during a task seems to be replaceable by the self-attention mechanism in transformer circuits; thus, there may be similarities between the human brain and transformers in the mechanisms underlying their adaptability to novel tasks. Understanding human mechanisms requires studies on the human brain, but strict restrictions on invasive experiments mean that it is difficult to obtain the neuronal activity data needed to reveal circuit computations in the human brain. Therefore, creating AI circuits that imitate human brain circuits to replicate the adaptability to novel tasks is a promising approach. When creating artificial brain-imitating circuits, given the conditions under which GPT-3 acquired in-context learning, the scale of the circuit (i.e., the number of synapses) and the volume of learning data should be taken into account. This approach opens up many intriguing questions for exploration, such as whether the brain's adaptability to novel tasks involves the attention mechanism that switches the input/output transformation of the circuit depending on contextual information, and whether this adaptivity can be acquired solely through prediction-error learning.

**Action planning for robots utilizing sentence/video-generative AI**

Since the release of ChatGPT at the end of 2022, there has been a growing interest in methods for motor control (especially in robot control) that leverage generative AI, such as large-scale language models and video generative models [204]. In this method, when a robot receives a human command and a goal in text form, the generative model provides a sequence of actions in text or video suitable for achieving that goal.

As a typical example, Google's PaLM-SayCan describes the robot's situation and the problem and, in language, proposes strategies and procedures for solving the problem [205]. Among them, PaLM-SayCan determines the action sequence (procedure) of the robot. For example, if you ask the robot to "Help me, I spilled a drink on the table," PaLM-SayCan's large-scale language model PaLM (a transformer that can process both vision and language, discussed later) verbalizes the situation obtained from the camera of the robot and proposes candidates for the first action (e.g., "find the spilled drink"), then the controller SayCan evaluates the feasibility (affordance function) of each action, and executes the best action. The controller unit updates the feasibility prediction according to the success or failure of the action (success brings the feasibility closer to 100%; failure brings it closer to 0%). The process is then repeated to generate a sequence of actions. For this action planning, a video generative AI (e.g., a vision transformer, such as Stable Diffusion or DALL-E2) can provide candidates for the action sequence in video format (UniSim). Multiple methods have been proposed to use video generative AI for generation of direct action planning in robot control (UniPi; Google, MIT; RFM-1; GR-1), and for training robot control through imitation learning (TRI).

The mechanism by which robots use generative models for action planning also provides insights for brain research. The neocortex contains a network called the mirror neuron system (primarily located in higher-order motor association areas), which acquires models of actions by observing the actions of others [206-208]. Interestingly, it has been suggested that the properties of mirror neurons are acquired by predicting the actions that others will perform next [209], and this acquired mirror neuron system has been demonstrated to be directly involved in both action planning and imitation learning [208,210]. This parallels the fact that video-generating AI trained through prediction-error learning can be used for both action planning and imitation learning. Referring to the mechanisms by which generative AIs generate action sequences could provide clues to advance our understanding of the computational mechanisms by which the mirror neuron system is involved in planning of actions and action sequences in the brain.

**Further comparison between the neocortex and transformers: Links between information-processing sharing and advanced abstraction of processing**

One of the features of information processing in the powerfully adaptive and multipurpose neocortex is that a neocortical circuit (area) can be shared for quite different types of information processing. For instance, some language-processing and tool-use circuits are shared, and training in one function can enhance performance in the other [211,212]. The reason for this may be that the same abstracted information processing is hidden within the processing for language rules (i.e., grammar) and tool-use rules (termed "action grammar"; e.g., opening scissors, inserting paper, closing scissors, then repeat). This circuit sharing suggests that the brain processes information at a highly abstract level, enabling the same circuit to handle multiple tasks.

Here, we look at AI circuits from the perspective of circuit sharing for different types of information processing. GNMT has different dedicated circuits for different languages, with the encoder processing the first language and the decoder processing the second language. By contrast, GPTs eliminate language-specific circuits and instead accept multilingual input (GPTs simply treat "こんにちは," "Hello," and 'Hola' as different inputs), allowing a single circuit to be trained through multilingual word prediction. As a result, GPTs became capable of multilingual input/output from a common circuit, and their performance surpassed that of GNMT. Interestingly, in GPTs, knowledge and skills acquired through learning in one language are transferred to other languages, indicating that the central part of the information processing is shared among multiple languages. Similarly, in humans, research on the bilingual brain indicates that when a second language is learned in adulthood, the two languages are processed in separate circuits, whereas when two languages are learned in childhood, they are processed in a shared circuit, resulting in higher performance in the second language [213,214]. The latter characteristics are similar to those of GPTs, which process multiple languages in a single circuit and achieve higher performance. There are several potential advantages to multiple functions sharing a single circuit. First, it increases the total amount of learning data by allowing multiple functions to train the same circuit. Second, when different functions share abstracted similarities, learning for one function can improve the processing capabilities in the others, leading to more efficient learning. Third, shared processing requires a high degree of abstraction of processing to extract commonalities across processing tasks, which dramatically enhance versatility and adaptability to novel tasks.

The transformer circuit now has further enhanced input flexibility, with a single transformer circuit able to handle both language and visual processing. Vision transformers process images by splitting the image into puzzle-like pieces, with input neurons receiving individual pieces of information; by arranging the word input neurons alongside the image input neurons, multimodal transformers can receive both language and visual inputs (Gato, PaLM-E)[175,215]. At first glance, transformer circuits with multi-modal processing capability appear to be a development in a direction far away from brain processing. However, the brain contains areas called network hubs, such as the default mode network, that are involved in diverse cognitive processes, including language processing, visual recognition, and social cognition [216-219]. The similarity between brain network hubs and multimodal transformers suggests that the network hubs in the brain may be responsible for abstract processing common to diverse cognitive processes. Comparing AI with the brain will shed light on how the brain uses shared processing circuits to accomplish diverse and sophisticated information processing.

**Convergent evolution of large language AI and human language processing**

The evolution of large-scale language AI from GNMT to BERT and then GPT shows an intriguing convergence with human language-processing mechanisms. Initially, regarding learning methods, GNMT relied on supervised learning using human-translated texts. BERT then advanced to unsupervised prediction-error learning, predicting masked words within sentences to acquire a general-purpose language comprehension unit capable of handling various language tasks. GPT then pushed this evolution further by switching to next-word prediction, a seemingly minor adjustment that led to a dramatic leap, endowing GPT with sentence-generation abilities that BERT lacked. This is because the GPT next-word prediction mechanism could be transferred to sentence generation by repeating the output, whereas the BERT masked-word prediction mechanism was difficult to apply to sentence generation. Interestingly, this evolution was not inspired by, but rather influenced, theories of neocortical language processing. The success of GPTs has prompted a growing body of studies suggesting that next-word prediction contributes to language learning and processing in the neocortex, challenging the dominant theory of innate language [150,157-159,161].

Regarding input/output and circuit structures, the evolution from GNMT to BERT also involved a shift from LSTMs (RNNs) to transformers. This transition represents a move away from the brain-like format of sequential word input and the use of RNN circuits imitating loop connections in the brain. However, the transformer architecture does include a neocortex-inspired attention mechanism absent in LSTMs. Furthermore, as they scale up, transformer circuits have unexpectedly acquired immediate adaptability to novel tasks, much like humans. This suggests that, in terms of circuit computation, large-scale transformers are indeed more similar to the neocortex than LSTMs and may provide new insights into neocortical circuit computations and structure.

The unexpected convergence of AI toward brain-like systems with each significant performance leap is particularly intriguing. The self-attention mechanism of transformers is also a candidate for further convergent evolution, which could be demonstrated by future brain research. Such convergence is likely to continue to emerge and provide new insights into the brain's information processing—making it essential to track the ongoing evolution of AI.

## 2.2 Evaluation and improvement of action selection

**Efficient reinforcement learning leveraging world models of the neocortex and cerebellum**

In the brain, reinforcement learning via the basal ganglia–dopaminergic system plays an essential role in optimal action selection [46,220,221]. Since this review focuses on the neocortex and cerebellum, we will only briefly touch on reinforcement learning in the basal ganglia–dopaminergic system and instead delve deeply into model-based reinforcement learning, which is related to the model functions of the neocortex and cerebellum.

In reinforcement learning, learning is based on the rewarded (or unrewarded) outcome of actions (unlike supervised or unsupervised learning). The mechanisms of reinforcement learning in the brain can be well described by the theory of reinforcement learning in machine learning, which has been developed to implement brain-like reinforcement learning in robots. In machine reinforcement learning, the temporal difference (TD) error signal is a quantified value indicating how much each action has improved the situation. The TD error signal is used to evaluate the action (i.e., to decide whether to repeat the action or not), which is the most important signal in reinforcement learning, and to update the evaluated value of the current state/situation, which is needed for updated calculation of the TD error signal. In the brain, the activity of dopamine neurons is thought to be equivalent to the TD error signal (Figure 5a,b) [46,220-224].

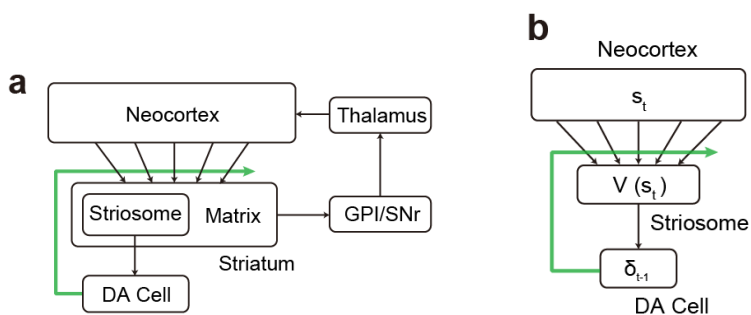

**Figure 5. Action selection by reinforcement learning in the brain and AI.**

**a**, Anatomical connections of neocortico-basal ganglia-dopamine (DA) neurons. The dopaminergic signals change the synaptic weights of neocortico-striatal projections (green arrow). **b**, Theoretical correspondence between the neocortico-basal ganglia-dopamine circuit and the reinforcement-learning computation. δ indicates the TD signal, V indicates the evaluated value of the current state/situation (termed the state value), and s indicates the neural representation of the current state/situation, S, in the environment. The neocortex can be viewed as a deep-layered circuit and other areas, such as the striatum and dopamine neurons, as single layers.

Reinforcement learning requires the input of an accurate neural representation s of the current environmental state S, which is provided primarily by the neocortex (Figure 5a,b). Standard reinforcement learning is referred to as "model-free reinforcement learning" because transition information between states S in the environment—that is, a model of the environment—is not available. In this case, the only way to fill in the evaluation values V(s) of all states is to experience (visit) all states. On the other hand, if complete transition information between states S is available (e.g., S2 transitions 100% to S3), then once the evaluation values V(s) of just a small fraction of the states have been obtained, the values of the remaining states can be calculated in a chain (e.g., if S3 has high value, S2 also has high value). The theoretical calculation formula for this chain is known as the Bellman equation, and it makes learning overwhelmingly faster because the evaluation values for unexperienced states become available. This is called "model-based reinforcement learning" because the learning is based on transition information between states S in the environment—that is, on the model of the environment in the brain. It has been demonstrated that the human brain performs this model-based state evaluation (Figure 5c)[31,225-228], with the models provided by cognitive regions in the neocortex and cerebellum (specifically, the prefrontal and parietal neocortices and the lateral part of the cerebellum) [31,33,229,230].

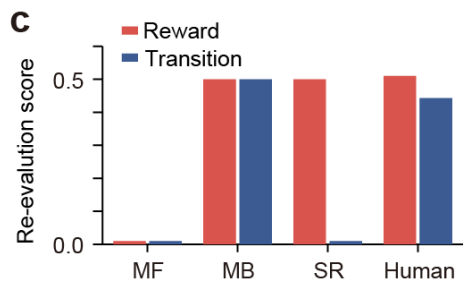

**Figure 5c**. Comparison of abilities between model-free reinforcement learning (MF), model-based reinforcement learning (MB), successor representation reinforcement learning (SR), and the human brain. Initially, agents experience all of the state transitions in an environment, including the link from the final state to the reward. Later, in the reward re-evaluation condition (red), the link from the final state to the reward is changed, the agents are allowed to experience only the changed transition, and then they are asked to evaluate the unexperienced relationship between the initial state and the reward. In the transition re-evaluation condition (blue), a state transition in the environment is changed, the agents are allowed to experience only the changed transition, and then they are asked to evaluate the unexperienced relationship between the initial state and the reward. The performance of the human brain closely matched that of the model-based reinforcement learning (although the efficiency in transition re-evaluation was slightly lower than that in reward re-evaluation), suggesting the capability of highly efficient reinforcement learning based on world models. Panel **c** adapted from Momennejad I. et al. (2017) with permission from Springer Nature, with reference to Vitay J. (2019).

**Deep and world-model-based reinforcement learning in AI**

In search of AI reinforcement-learning circuits with similarities to those in the brain, we here trace the history of deep reinforcement learning AI. In 2015, three years after the rise of deep learning by AlexNet, DeepMind created the first deep reinforcement learning circuit, Deep Q Network (DQN), combining a deep layer network and reinforcement learning to achieve cutting-edge video-game operation (Figure 5d) [64,231,232]. A feature of DQN is that it outputs the action value Q of reinforcement learning (the estimated value similar to V in Figure 5b, but for each action a in addition to the current state S), enabling it to play the game by selecting the action estimated to be of highest value. The input side of DQN is a CNN, responsible for video processing. Before DQN, the screen states of video games, S, were so enormous that they completely exceeded the range of what could be learned by trial and error in reinforcement learning, but the compression of screen states via the CNN enables a compact neural representation, s, of the states and hence reinforcement learning. As a result, DQN achieved human-level performance, and its successors, deep reinforcement learning circuits, continued to improve their skills, surpassing human performance in all Atari video games by 2020.

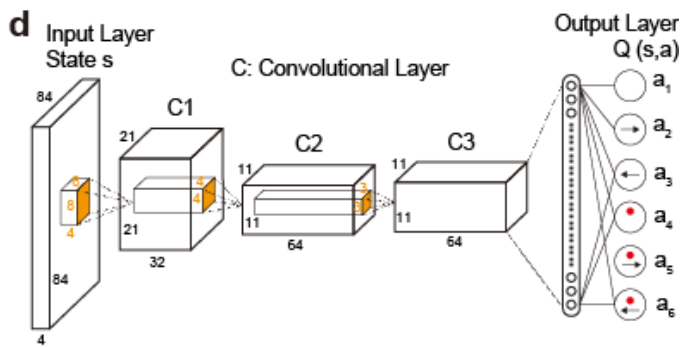

**Figure 5d**. Circuit diagram of the game-playing AI, Deep Q network (DQN)**.** Circuit structure: DQN consists of three layers of CNNs (C1-C3; convolutional filters are shown in orange) and two layers of fully connected neural networks. Input/output: The input is the pixel information from the four most recent frames of the video game, and the outputs are the estimated action values corresponding to, in this example, six button operations (for the Space Invaders game). The action with the highest estimated value (i.e., corresponding to the output neuron with the biggest signal) is selected as the next button operation. Note that DQN has limited access to information on past states and actions (even information on the most recent actions and fired shots is available only via the last four input frames).
Learning: The action value Q from reinforcement learning theory (the estimated value of taking an action, a, under the current state S) is calculated from the reward, which is defined as the increase in the game score. Since the output of the network is the estimated Q values for the six actions, all layers learn through backpropagation of the reinforcement learning signal to improve the estimated Q (i.e., to reduce the estimation error) and to maximize the reward. Panel **d** adapted with permission from Kitazawa S. (2020).

However, DQN (and its successors) has three major weaknesses [20,64,112,113]. First, learning in DQN is slow. DQN is a feed-forward circuit and does not process transitions between states (between frames). Thus, learning in DQN is model-free and hence slow [231]. (The DQN successor AlphaGo Zero is classified as using model-based reinforcement learning, but still inherits this model-free architecture; Supplementary Fig. 3) [64]. Second, the video-processing CNN part of DQN cannot be easily scaled up. In DQN, the only learning signal is a reinforcement learning signal (i.e., the estimation error of the action value), and even the CNN part of the circuit is trained by this learning signal, delivered by backpropagation. Reinforcement learning is reward-based learning, and among the vast number of states, usually only a very limited number of states are rewarded; that is, the rewards are sparse (e.g., in the game of GO, the reward of winning or losing occurs only at the end of the game, after hundreds of stone-placing actions). Sparse rewards make it difficult to train the large number of synapses in a large circuit, and so DQN has to be small-scale ("deep" but actually shallow layered) [112,233]. Third, DQN does not acquire versatility. Since even the sensory CNN part is trained by the reinforcement learning signal, it becomes specialized in extracting action/reward-related features. Because of this and the small capacity of the entire circuit, for DQN to be able to play new games, it is necessary to prepare a new DQN circuit and have it learn the new game from scratch through a large number of trials.

To overcome these weaknesses, model-based deep reinforcement learning circuits were developed, drawing inspiration from brain reinforcement learning. A representative example is the "world model" circuit, which divides a conventional DQN-type integrated circuit from sensory input to output into a large sensory-processing unit and a small action-control unit (shown in Figure 2e; here discussed from the viewpoint of reinforcement learning)[112,113]. The sensory processing unit consists of a CNN-RNN, with autoencoder learning of the CNN and prediction-error learning of the RNN, enabling training of a large, general-purpose sensory-processing unit. In particular, the RNN part was designed to acquire world models by learning causal relationships and transition rules between environmental states (between frames and events) through prediction-error learning, and to provide model-based information to the control unit. In fact, the world model circuit achieved more efficient learning than model-free reinforcement learning circuits and even succeeded in learning "in a dream": Concretely, external game input to the sensory-processing/world-model unit was stopped, and then the control unit was trained within the virtual game input predicted/generated by the world model. Later, when external input was resumed, the performance of the control unit was found to have improved (the successor circuit was named "Dreamer" to emphasize this functionality) [112,113]. In the world model circuit, the propagation of learning to inexperienced states is not through the mathematical Bellman equation, but through simulation of inexperienced states using the world models (i.e., virtual experience). Interestingly, it has been suggested that the propagation of learning to inexperienced situations in humans is also mediated by simulations that use world models [31], suggesting similarity with world model AI.

The CNN-RNN unit of the world model circuit, which acquires world models through prediction-error learning and provides model information to the reinforcement-learning control unit, has similarities in terms of circuit structure, input/output, and learning mechanism to the neocortex and the cerebellum (see Section 1.2), which provide model information to the reinforcement learning module of the brain (i.e., the basal ganglia-dopaminergic system). By creating artificial neural circuits that precisely imitate the brain, we will gain new insights into the mechanisms of model-based reinforcement learning in the brain, such as what kind of model information the reinforcement learning module receives in the human brain, and how humans learn from simulations of inexperienced situations.

## 2.3. Motor command generation: Forward and inverse internal models

In human and robotic motor control, motor commands—dynamic signals for movement—need to be generated and sent to the actuators (muscles or motors) to accelerate and decelerate the movement. The motor neocortex is essential for the generation of dynamic signals suitable for movements and output of motor commands: its damage causes motor paralysis [234-243]. However, signals from the cerebellum are also necessary particularly to make these motor commands highly accurate and sophisticated; damage to the cerebellum results not only in loss of motor accuracy (cerebellar ataxia), but also in severe impairment of motor learning (adaptation deficits) [14,244]. This section focuses on the mechanisms for generating highly precise motor commands. The precise motor control enabled by the cerebellum is fundamentally based on acquiring and utilizing world models (including models of the body and muscles), called cerebellar internal models [14,36-43].

We here outline why internal models are needed to generate motor commands in the brain and compare this mechanism with AI command generation. For both brain and AI motor control, the total duration of typical movements ranges from a few tens of milliseconds to 200 milliseconds, and motor commands must be dynamic, adjusting in time steps of about 5 to 10 milliseconds. But, the constraints of the objects controlled by AI and by the brain—that is, a robot versus an animal body—are very different. Starting with the one with fewer constraints, the motor-controlled objects of AI are endowed with excellent sensors and actuators with sub-millisecond time responsiveness. Therefore, AI can implement standard engineering feedback control, in which a motor command is sent at each time step to drive the actuators, and at the next time step, feedback from the sensors is available to determine the next command. Such cycles, in which an action (command) is determined and executed in a given world state, updated state information becomes available, and the next action is determined, are called Markov decision processes and can be handled within the framework of reinforcement learning; thus, the dominant motor control in AI is feedback control optimized through reinforcement learning.

Animal bodies, on the other hand, are equipped with sensors and actuators (muscles) with a significant time delay: there is a minimum delay of approximately 100–200 milliseconds from sensor pickup of the signal to muscle contraction. (For example, there is a 40-millisecond delay in the retina alone, plus a 10–20 millisecond delay for the muscle to respond to a motor commands, plus signal transmission time) [245,246]. Because of these constraints, the brain cannot simply determine the next command on the basis of feedback on the result of the most recent command (the movement would be completed while waiting for such feedback). In the past, it was proposed that even without feedback control, movement might be controlled through simple calculations using the mechanical properties of muscles (such as the equilibrium-point control hypothesis), but this possibility was later ruled out [247]. Now, it is thought that the brain prepares the entire set of motor commands before the movement begins. To accomplish this, various predictive functions are crucial, and thus the brain relies on the predictive capabilities of the internal models acquired by the cerebellum [14,36-43].

Because there are significant differences in the motor control strategies enacted by the brain and by AI, and because research on both is still ongoing, there is not yet a close relationship between the brain and AI for motor control, as there is for sensory processing. AIs for motor control cannot employ CNNs or supervised learning, both of which have been successfully implemented in visual processing (note that CNNs cannot generate dynamic signals, and supervised signals for motor commands cannot be prepared manually). AI optimizes feedback control through reinforcement learning, but it still cannot match the precise motor control and rapid learning capability of humans (e.g., DARPA robotic challenge 2015)[20,248]. In this context, mechanisms of motor control in the brain have implications for the future development of motor-control AI. With this in mind, we next describe the generation of motor commands mediated by cerebellar internal models, and compare it with command-generation in AI.

**Generation of motor commands by the cerebellar forward internal model**

The brain compensates for the time delay in motor control by using the predictive capability of the cerebellar internal models to generate an entire sequence of motor commands from the beginning to the end of the movement without feedback [37,38,41]. With cerebellar damage, the movement itself can still be carried out but it is less precise (e.g., inability to decelerate properly results in overshooting and moving back and forth around a target). Because of this, the cerebellum is thought to be responsible for refining the rough motor commands generated by command centers outside the cerebellum (i.e., motor neocortex, primitive motor control systems in the spinal cord, brainstem, etc.). Since the rough motor command generated by the extracerebellar command centers would result in discrepancies between the generated and ideal movements, it must be corrected swiftly without reliance on slow biological feedback. For this purpose, the cerebellar internal model predicts the future motor outcome (represented by sensory signals) on the basis of the current world state and the copy of motor command for muscles, and provides this prediction to the extracerebellar command generator (Figure 6a). This is the most basic theory of motor control by the cerebellar internal model [39,40,246,249]. This internal model is called the forward internal model (or, more simply, the forward model), as it is a model of the world that follows the causal order from the cause (motor command) to the outcome (sensory information). In other words, the role of the forward model is to generate a virtual feedback signal, because the actual feedback is too slow in biological systems.

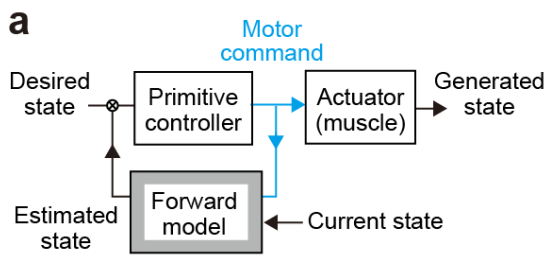

**Figure 6. Precise motor control mediated by cerebellar internal models.**

a. Block diagram of motor control by the forward internal model of the cerebellum. Note that although this is a motor control system, the main role of the internal model (cerebellum, gray) is motor–sensory transformation, predicting and outputting future sensory signals (for the future state, including the motor outcome) on the basis of the copy of the motor command for muscles (blue) and the current sensory signal (for the current state). Panel **a** adapted from Miall R.C. et al. (1993) with permission from Taylor & Francis Ltd.

The predictive function of the cerebellar internal model can be embedded into a three-layer RNN circuit imitating the cerebellum (Figure 3a). This three-layer RNN circuit takes as input the current sensory signal and the motor command from the extracerebellar command generator, and generates as output the future sensory signal. Upon receiving the actual sensory signal at the next time step, the inferior olive calculates the prediction error to improve future cerebellar predictions. Through this prediction-error learning, the three-layer RNN circuit learns how the motor command interferes with the current state to produce the next state, and models the causal relationship from the motor command (as cause) to the movement (as result). This modeling of the world and causal relationships is almost identical to the world models acquired in neocortical sensory processing. Indeed, it has been proposed that the predictive coding circuit (a deep-layered prediction-error learning RNN and a theoretical circuit for visual processing in the neocortex) can realize motor control by utilizing its world modeling function [12,29,103]. This hypothesis suggests that, as in sensory processing, RNN circuits that have acquired world models through prediction-error learning play a central role in motor processing as well.

**Generation of motor commands by the cerebellar inverse internal model**

However, motor command generation using a forward internal model (where the cerebellum only indirectly contributes to motor command generation) is far from a complete control system. First, because the final motor command in this system is generated by adding a correction signal to the inaccurate extracerebellar motor command, the system is required to continuously generate the correction signal even after the movement becomes proficient (e.g., if the extracerebellar motor command is too strong, the cerebellum continues to make predictions of passing the target even after learning, and the extracerebellar command generator must add a correction command). Second, this system does not easily allow for refinement of the motor command. For instance, as a movement becomes proficient, not only does the endpoint error decrease, but a smooth and efficient movement is acquired, which is achieved by optimizing the motor command (avoiding large signal intensities and abrupt changes in the command) [250,251]. The cerebellum, which is central to motor learning, is also responsible for optimizing motor commands. However, given that the role of the forward internal model is to predict the outcome of motor commands, optimizing and improving the motor commands themselves is difficult for the forward model, since it only indirectly contributes to motor commands via the extracerebellar command generator.

Furthermore, altering an output of the forward internal model to improve the motor command could degrade its primary function of predicting the motor outcome.

For these reasons, it is believed that there is another mechanism that allows the brain to learn motor commands directly in order to generate accurate and flexible motor commands. The proposed mechanism for this is termed the inverse internal model (or inverse model)[37,41,43]. The function of this model is to inversely estimate the motor command (cause) from the desired motor outcome (represented by sensory signals), in contrast to the forward model that performs a cause-to-outcome transformation. While several mechanisms have been proposed for acquisition of the inverse model, the one that is the most biologically feasible and that has accumulated physiological evidence is learning to minimizing the extracerebellar motor command (Figure 6b; this is traditionally called feedback-error learning, but we avoid this terminology to minimize confusion with feedback control)[38-40,252-255]. The ingenious point of this design is that the inverse model (in the cerebellum) and the extracerebellar command generator are arranged in parallel, and the signals from the extracerebellar command generator are treated as error signals, not as prediction targets/correct signals [38,39]. If the extracerebellar signal was a prediction target, the inverse model would simply learn to copy the signal from the extracerebellar command generator. Instead, by treating it as an error signal, the inverse model learns to minimize this error and performs a predictive control role so that the extracerebellar command generator has to do as little work as possible. In other words, if the output of the extracerebellar command generator is zero during and after the inverse model generates a command, it indicates that the inverse model has generated a motor command that requires no correction. Despite some confusion even among neuroscientists, the forward and inverse internal models are not opposing concepts, but rather, the combination of the two results in a more powerful motor command generator, for which theoretical and physiological evidence is accumulating (for a review, see Kawato, Ohmae, et al. 2021)[37,40,111,256,257]. Learning by the inverse internal model can also be considered prediction-error learning (in a broad sense), whereby the signal generated by the extracerebellar command generator is treated as a prediction error (i.e., prediction failure). Therefore, we can use a three-layer RNN circuit to imitate the cerebellum and acquire the inverse internal model, to investigate the motor command generation by the cerebellum.

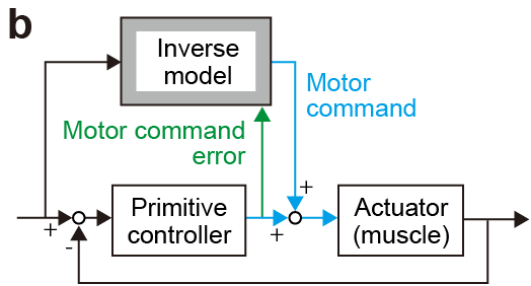

**Figure 6b**, Block diagram of precise motor control mediated by the cerebellar inverse internal model (inverse model). The inverse model (cerebellum, gray) sends motor commands (blue) to the actuator (muscle) in parallel with the extracerebellar motor command generator and learns to minimize the extracerebellar output as an error signal (green), thereby acquiring sensorimotor transformation. As an example, in saccadic adaptation (e.g., when the eye is moved to the right to capture a point on the right side of the visual field in the center of the retina, if an experimenter shifts the target point slightly to the left during eye movement, initially, a full-size error occurs, but as this is repeated, the error is reduced as the shift becomes incorporated and the rightward movement becomes smaller), during the early stages of this adaptation, the superior colliculus, a primitive motor command generator, generates the error-correcting eye movements, and a copy of this motor command is sent as an error signal to the cerebellum via the inferior olive. The cerebellum learns to reduce this error signal, that is, to reduce the output of the superior colliculus, resulting in reduced (adapted) eye movements (Kojima Y. & Soetedjo R., 2018; Herzfeld D. et al. 2018). This system can be interpreted as the inverse model learning to minimize the extracerebellar motor command. (For more evidence of the inverse model, refer to the following review articles: Kawato M. 1999; Kawato M., Ohmae S., et al. 2022). This input–output relationship and learning can also be embedded in a prediction-error-learning RNN circuit. Panel **b** adapted from Wolpert, Miall, & Kawato (1998) with permission from Elsevier.

**Hints for improving AI motor control**

Motor control of robots by AI is quite different from motor control by the brain mediated by internal models. As mentioned above, the major strategy for optimizing AI motor control is reinforcement learning. For example, in 2016, Google's DeepMind used shared training data from 14 robotic arms in parallel to train a deep reinforcement learning controller in a variety of grasping actions [258]. In 2017, the same company successfully controlled avatar walking in virtual space through the use of deep reinforcement learning [259]. Furthermore, in 2023, they trained a single transformer circuit as a general-purpose generative model across language, video, and even video-game/robotic action (using the circuit-sharing method described above; Gato)[175], and improved the actions generated by the transformer circuit through reinforcement learning, thereby creating a controller that could relatively quickly learn new tasks (RoboCat)[176].

However, despite the superior performance of sensors and actuators, current AI motor control still lags behind humans in terms of precision and learning speed (DARPA robotic challenge)[20,248]. Since AI learning requires a large number of trials, learning is often done in virtual space, but such virtual learning is often not transferable to the real world (hence, how to transfer from simulation to reality—called Sim2real—is a hot topic). In general, reinforcement learning has the disadvantage of having limited information in the learning signal. Reward information is sparse in time and is only one-dimensional information (about whether the reward is large or small). On the other hand, in prediction-error learning, learning signals are generated at every time point, and the amount of information is very large (e.g., 10,000-dimensional information from a 100×100-pixel image)[20,112,113,233]. Therefore, to generate motor commands through information-rich prediction-error learning, AI motor control can be inspired by motor control in the brain, which employs forward and inverse internal models. For example, if we trained AI to acquire an inverse model by treating the current best controller as a primitive controller in Figure 6b providing the error for the model, we would expect to see a dramatic improvement in AI-based motor control, as well as a better understanding of the circuit computation mechanisms of the cerebellum.

# Discussion

**Proposal for a new theory of neuroscience.**

We have compared neocortical, cerebellar, and AI circuits across sensory, cognitive, and motor domains from the unified viewpoints of circuit structure, input/output signals, and learning, highlighting notable similarities between the brain and AI systems. Our comprehensive comparison revealed that large-scale AI circuits that acquire world models through prediction-error learning (in particular, those after BERT) are converging with the brain in terms of computations, input/outputs, and learning. Drawing on the similarities, we have developed the following new theory on the universal information processing mechanisms of the neocortex and cerebellum.

Regarding input/outputs and learning: The neocortex and cerebellum predict future states of the external world from past inputs and learn to minimize prediction error. Through this process, the neocortex and cerebellum compress and abstract information about the external world to form compact, powerful world models, and by utilizing the models, they realize three fundamental types of information processing. (1) Prediction: Generating future information. (2) Understanding: Interpreting the external world by using abstract information within the world models, such as laws and causal relationships (e.g., object recognition and text comprehension). (3) Generation: Repurposing the future-information generation mechanism (i.e., the prediction function) to generate other types of information output (e.g., action planning, language planning, and imitation learning). The brain combines these three fundamental types of information processing to accomplish sensory, cognitive, and motor processing. This is the secret that allows the neocortex and cerebellum to accomplish a diverse range of functions despite their relatively uniform circuit structures and computations. The proposed theory is an integrative extension of existing neuroscience theories, inspired by AI processing. Specifically, points (1) to (3) are grounded in established neuroscience theories: Point (1) aligns with the internal model theory (see sections 1.1, 1.2, 2.2, and 2.3), point (2) with sensory processing theory (see sections 1.1 and 1.2), and point (3) with the mirror neuron system theory (see section 2.1, particularly for language and action planning). Traditionally, neuroscience treated these as largely separate theories to explain different brain functions in different brain regions. However, in large-scale model-acquiring AI, the prediction circuit that is learned through prediction-error learning is directly repurposed for understanding and generation, integrating all three functions within a single circuit computation. Inspired by this, we propose a unified theory in which the neocortex and cerebellum similarly integrate the three functions, applying these processes universally across sensory, cognitive, and motor domains. Experimental studies support this theory: Research indicates that next-word prediction circuits are also involved in sentence generation in the neocortex and cerebellum; other research indicates that the neocortical mirror neuron system develops through observation and prediction of others' actions and is also used to generate self-action plans (see section 2.1).

     Next, regarding circuit structure: Currently, although no AI circuit yet fully imitates the neocortical circuitry, there are plenty of similarities in information processing between the neocortex and AI circuits trained by prediction-error learning. Deep-layered autoencoder and CNN circuits resemble the neocortex in processing static input that does not require temporal integration. Meanwhile, deep-layered predictive coding networks, CNN-RNNs, deep-layered LSTMs, and deep-layered transformer circuits share similarities with the neocortex in their ability to process dynamic inputs that do require temporal integration. Notably, although transformer circuits do not imitate the neocortex in circuit structure, their use of a neocortex-like attentional mechanism has enabled them not only to display versatility—achieving human-like accuracy in a wide range of processing domains, including vision and language—but also to acquire the human-characteristic ability to quickly adapt to new tasks. These indicate multiple intriguing similarities between transformers and the neocortex (see Sections 1.2 and 2.1). The structure and local circuit computations of the neocortex are still not fully understood. A reverse-engineering approach that explores the circuit structures

necessary to implement transformer-like attention mechanisms within biological neocortical circuits could yield insights worthy of experimental validation.

The cerebellum has often been described as a feedforward circuit, but recent evidence indicates that it functions as an RNN circuit, and recurrent computations are essential for the general-purpose and powerful capabilities of the cerebellar circuit (see Sections 1.2, 2.2, and 2.3). The cerebellum shares similar learning and input–output characteristics of internal world models with the neocortex but is approximated by a simpler three-layer RNN circuit. Since the cerebellum contains approximately four times as many cells as the neocortex [260-262], it can be considered a large-scale RNN comparable to the neocortex in terms of computational scale. Although the neocortex and cerebellum have similar input–output characteristics, their different circuit architectures may lead to different strengths (for example, the neocortex, with its attention mechanism, is capable of advanced processing, whereas the cerebellum enables rapid and efficient learning and computation). The neocortex and cerebellum cooperate, and this cooperation mechanism likely combines their strengths to achieve both advanced and rapid processing [13,116,263].

**Unsupervised prediction-error learning and world models in the neocortex and cerebellum**

Our proposed perspective that the neocortex and cerebellum acquire world models through unsupervised prediction-error learning provides an essential new viewpoint for interpreting the signals and processing observed in these regions. Psychological tasks such as operant conditioning are far simpler than most tasks in the real world (because the number of possible states/situations is so limited that all can be experienced) and are solvable by classical model-free reinforcement learning alone. As a result, it has been implicitly assumed that the signals observed in the neocortex and cerebellum during these tasks are signals related to classical reinforcement learning without a world model. For example, reward signals found in these regions have been interpreted as learning signals for improving behavior in model-free reinforcement learning [264-272]. However, these regions are also able to acquire world models through prediction-error learning, and the human brain has capability for model-based reinforcement learning that efficiently utilizes world models (Figure 5c). From the perspective of world models, the reward even in operant conditioning is understood as just one important event in the world, and the reward signal is a signal associated with prediction-error learning of the reward event (e.g., the correct signal for predicting the event), or the representation of the reward event in the world model (i.e., the signal to predict the event or the abstracted representation of it). In other words, the reward signal in the brain is not necessarily limited to the learning signal of classical model-free reinforcement learning. In future research on the neocortex and cerebellum, the perspective of unsupervised learning of world models will lead to more careful and accurate interpretation of signals in the brain.

**Significance of scaling up the circuit size**

Recent advances in AI have been remarkable, particularly since the success of CNNs. The foundational structures of CNN and RNN (LSTM) circuits were established long ago [53,156,273,274], and a key factor behind the breakthroughs was the scaling up of both the circuit size and training data [275-277]. The conventional theory of AI has argued that although increasing the size of training data could be beneficial, excessively scaling up the circuit would result in too much capacity: this would lead to overfitting whereby the circuit memorizes non-essential features of individual training data, thereby reducing its ability to generalize to new data. However, in practice, large-scale AI circuits did not lead to overfitting. On the contrary, scaling up transformer circuits resulted in the emergence of capabilities such as in-context learning, similar to the immediate adaptability to novel tasks seen in humans. Given the similarities between transformers and the neocortex, how a circuit with a vast number of synapses is capable of essential learning without overfitting and of emergently acquiring adaptability to new tasks remains an open question with implications for neuroscience, since the brain has more than 100 trillion synapses (larger than ChatGPT4 or the largest AI circuit in existence today).

      Research on this mechanism in AI has only just begun, and there is no conclusive theory yet, but we here outline relevant discoveries and representative hypotheses. First, the artificial neural circuit contains numerous subnetworks, and as the circuit is scaled up, the number of the subnetworks increases exponentially (Figure 7). During circuit learning, a subnetwork whose initial values happen to be well-suited for information processing contributes to the desired output and is reinforced, eventually playing a major role in information processing of the entire circuit. Therefore, the large circuit is more likely to contain such a "hit" sub-network and to be successfully trained. This idea was named the "lottery ticket hypothesis" and has experimental support [278,279]. This hypothesis implies two points: (1) Even if a large circuit is being trained, only a small fraction of the circuit actually contributes to the output after learning (This is called "implicit regularization" because it is as if a rule that facilitates learning in only a small fraction of the synapses has been added). (2) By scaling up the circuit, the number of subnetworks increases sharply, leading to an enhanced probability that the circuit will contain a subnetwork with complex processing capabilities. In line with (1), it was found that only a small portion of the vast number of synapses in the large AI circuit are used after training, and that even if 90% or more of weak synapses are pruned (i.e., further reduced to zero), most of the performance gained from learning can be maintained [280-

[283]. This in turn means that a large-scale circuit is necessary only during the learning phase [278,279]. Point (2) suggests that when learning sophisticated information processing like in-context learning, which requires complex processing mechanisms (see "Immediate Adaptability to Novel Tasks" in motor language system)[200,201], the existence of a pre-learning subnetwork that happens to be well-suited for such complex processing beforehand is essential for successful learning.

In the brain, related to point (1), weak synapses are pruned, and the associated molecular mechanisms have been identified (e.g., Arc's synapto-tagging)[284]. Consistent with point (2), there is already considerable similarity between the language signals in a pre-learning transformer circuit and those in the neocortex [158]. We are in an era where we can utilize interdisciplinary approaches involving brain-imitating AI circuits, computational neuroscience, and molecular biology methods to explore why the large-scale circuit in the brain does not overfit and instead acquires advanced intelligence, with particular attention to the subnetwork structure that contributes to the learning process.

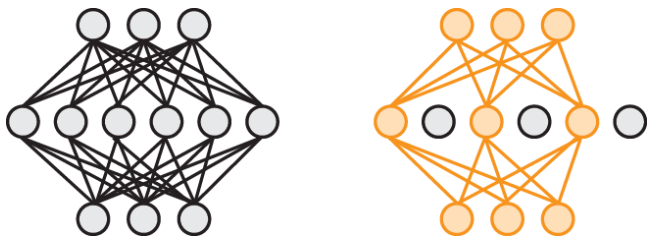

**Figure 7. Lottery ticket hypothesis.** This hypothesis about neural network learning suggests that a pre-learning subnetwork that happens to have appropriate initial values suitable for learning continues to be reinforced and eventually becomes responsible for the information processing of the entire circuit. For example, in a three-layer 3-6-3 network the number of 3-3-3 subnetworks is Combination(6, 3) = 20; as the circuit scale increases, the number of subnetworks of a particular size increases exponentially. These subnetworks overlap, but no pair has exactly the same initial value combination.

**The importance of circuit scale in neuroscience**

The fact that large-scale circuitry in AI was a key factor in acquiring advanced processing capabilities had a profound impact on AI research and computational neuroscience, where researchers had been focusing on designing intelligent circuits utilizing sophisticated information theory and competing to improve their performance (Sutton, "The Bitter Success")[277]. This finding has two implications for future neuroscience research.

First, with the relative ease of creating artificial circuits imitating the brain, neuroscience research combining experiments and artificial neural networks will become increasingly prominent. Conventional circuits designed by computational neuroscientists were based on sophisticated theories and were often esoteric to other neuroscientists, but the advent of CNNs and transformers has made it possible to compare the brain with simpler (albeit large-scale) artificial neural circuits. As a result, the use of artificial neural networks is becoming an essential approach across neuroscience research. That said, the information processing in AI circuits remains a black box, and it is largely unclear what specific steps and intermediate stages are taken to process information. As we move forward, the rich reservoir of neuroscience theories on computational mechanisms with sophisticated designs will play a crucial role in understanding the internal processing of AI circuits and brain-like circuits.

Second, the emergence of immediate adaptability to novel tasks through the scaling up of circuits and training data in AI will raise awareness of the importance of circuit scale in neuroscience research. For example, this AI finding suggests that to elucidate sophisticated human-characteristic processing, human brain circuits, which are far larger than those of mice or monkeys, need to be examined. On the other hand, to elucidate the design principles of brain circuits, it is beneficial to study non-human brains, which are similar in design to the human brain, albeit on a smaller scale. Although the view that intelligence lies not in cleverly designed circuits but in large circuits has not been given much weight in neuroscience, awareness of the scale of circuits will be essential for future research on human intelligence.

**Leveraging AI insights to understand neurological disorders**

Insights from AI research can deepen our understanding not only of the brain, but also of neurological disorders. As an example, here we discuss how the learning processes of AI circuits can provide insight into the mechanisms of learning in the brain and the onset of neurodevelopmental disorders. One of the major mysteries of neurodevelopmental disorders is that their onset is probabilistic and not always reproducible. In autism, for instance, despite the hundreds of genetic, environmental, and non-genetic risk factors, none cause autism 100% of time, though all increase the probability of onset [285,286]. While the interaction of multiple risk factors is considered critical [286,287], a decisive combination of interactions has yet to be identified.

By analogy with the fact that learning in AI circuits is also probabilistic (depending on initial values, etc.), neurodevelopmental disorders may be viewed as probabilistic impairments in learning. When modern AI circuits are trained, data of synaptic weights is stored as the training progresses, allowing for a restart from just before the point of learning failure, when needed. In addition, multiple circuits are typically trained in parallel, and if any one of them succeeds, the project is completed. For these reasons, the success rate of AI circuit learning is not usually a major concern, but it is known that scaling up the circuit size, designing the circuit architecture appropriately, and selecting an appropriate learning method can stabilize learning and improve the success rate. From these findings in AI research, we infer that in neurodevelopmental disorders, genetic and non-genetic risk factors destabilize learning and, in unfortunate cases, lead to failure to achieve the appropriate developmental goal and onset of a disorder. In this sense, learning stability may be key to understanding the mechanisms underlying neurodevelopmental disorders, but there are no traditional methods to measure and assess this stability. To fill this gap, we can create brain-imitating circuits and introduce manipulations to simulate the risk factors, allowing us to assess learning stability in the developing brain. Thus, by drawing insights from AI learning and utilizing brain-imitating AI, our understanding of the onset mechanisms of neurodevelopmental disorders could advance significantly.

**Directions for future neuroscience research**

Finally, we propose some general directions for future neuroscience research. To date, neuroscience research on brain functions focused on accurately describing the various functions of the brain with appropriate linguistic words. While this approach has significantly advanced our understanding of the brain, it contains weaknesses, including overlooking functions that are difficult to capture in words and emphasizing only limited aspects of complex processing [2].

First, the conventional approach of finding neural activity that matches a conceptual image in language does not work well when the actual neural representation is complex, abstract, and very different from the linguistic concept. For example, a neural representation of 3D space based on 2D visual images is essential for visual processing in the brain, but a representation that clearly matches the term "neural representation of 3D space" has not yet been reported. Similarly, AI can acquire a 3D internal representation (Figure 2g,h), but the internal representation is more abstract than those previously proposed in the computer science field[115]. Understanding abstract neural representations is challenging, but rather than just isolating the parts that are easy to understand in words, we need to comprehend the full principles of the abstract representations by comparing them with internal representations of AI.

In addition, when brain circuits process information in complex and abstract ways, the conventional approach tends to break this processing down into multiple functions that can be linguistically described. For instance, there has long been debate about the function of the parietal neocortex, with one view being that it provides attention that enhances certain sensory signals versus another that it encodes movement intention in response to certain sensory signals; however, recent studies have led to the conclusion that both elements exist [288,289]. Given that in recent AI transformer circuits (such as GPT), the functions of attention and text generation—which could serve as intention for robot motor control—cannot be separated, it is possible that, similarly, in the parietal lobe of the brain, the functions of attention and intention do not simply coexist but are different aspects of processing within the single circuit.

Furthermore, to date, prefrontal neocortex appears to have a wide range of functions, including short-term memory, spatial memory, executive function, decision-making, motivation, planning, switching, impulse control, social cognition, and performance monitoring [290]. However, beyond simply understanding this region as a set of many functions distinguished by language, exploring the fundamental essence of prefrontal processing will be an important future direction for neuroscience. To achieve this, we should not only approach brain functions from multiple language-based perspectives, but also understand the full scope of input–output transformations and dynamics in the brain by creating brain-imitating AI circuits and by incorporating theoretical findings on AI circuit computations. This will be a highly powerful approach to neuroscience in an era where AI is accessible to everyone.

92   Softky, W. R. Fine analog coding minimizes information transmission. *Neural Networks* **9**, 15-24 (1996).
93   Wiskott, L. & Sejnowski, T. J. Slow feature analysis: unsupervised learning of invariances. *Neural Comput* **14**, 715-770, doi:10.1162/089976602317318938 (2002).
94   George, D. & Hawkins, J. A hierarchical Bayesian model of invariant pattern recognition in the visual cortex. Report No. 0780390482, (IEEE, Montreal, QC, Canada, 2005).
95   Palm, R. B. *Prediction as a candidate for learning deep hierarchical models of data* Master of Science thesis, Technical University of Denmark, (2012).
96   Goroshin, R., Bruna, J., Tompson, J., Eigen, D. & LeCun, Y. Unsupervised Learning of Spatiotemporally Coherent Metrics. Report No. 23807504, (IEEE, Santiago, Chile, 2015).
97   Mathieu, M. F. *et al.* Disentangling factors of variation in deep representations using adversarial training. *Advances in neural information processing systems* **29** (2016).
98   Srivastava, N., Mansimov, E. & Salakhudinov, R. Unsupervised Learning of Video Representations using LSTMs. 843--852 (PMLR, Proceedings of Machine Learning Research, 2015).
99   Wang, X. & Gupta, A. Unsupervised Learning of Visual Representations Using Videos. (IEEE, Santiago, Chile, 2015).
100  Whitney, W. F., Chang, M., Kulkarni, T. & Tenenbaum, J. B. Understanding Visual Concepts with Continuation Learning. (San Juan, Puerto Rico, 2016).
101  Agrawal, P., Carreira, J. & Malik, J. Learning to see by moving. 37-45 (2015).
102  Kawato, M., Hayakawa, H. & Inui, T. A forward-inverse optics model of reciprocal connections between visual cortical areas. *Network: Computation in Neural Systems* **4**, 415–422 (1993).
103  Jiang, L. P. & Rao, R. P. N. Dynamic predictive coding: A model of hierarchical sequence learning and prediction in the neocortex. *PLoS Comput Biol* **20**, e1011801, doi:10.1371/journal.pcbi.1011801 (2024).
104  Rao, R. P. & Ballard, D. H. Predictive coding in the visual cortex: a functional interpretation of some extra-classical receptive-field effects. *Nature neuroscience* **2**, 79-87 (1999).
105  Fiser, A. *et al.* Experience-dependent spatial expectations in mouse visual cortex. *Nat Neurosci* **19**, 1658-1664, doi:10.1038/nn.4385 (2016).
106  Jiang, Y. *et al.* Constructing the hierarchy of predictive auditory sequences in the marmoset brain. *eLife* (2022).
107  Larkum, M. A cellular mechanism for cortical associations: an organizing principle for the cerebral cortex. *Trends Neurosci* **36**, 141-151, doi:10.1016/j.tins.2012.11.006 (2013).
108  Auksztulewicz, R. & Friston, K. Repetition suppression and its contextual determinants in predictive coding. *Cortex* **80**, 125-140, doi:10.1016/j.cortex.2015.11.024 (2016).
109  Mikulasch, F. A., Rudelt, L., Wibral, M. & Priesemann, V. Dendritic predictive coding: A theory of cortical computation with spiking neurons. *arXiv preprint arXiv:2205.05303* (2022).
58

**ACKNOWLEDGMENTS**

The authors thank lab members Chun Zhao and Tuo Xin for various contributions, including figure and reference editing and translation assistance, and Ouni Cao for his help with reference searches. We also thank Mrs. Zoha Hassan for her meticulous editing, with attention to linguistic nuances. We extend our gratitude to Profs. Tatsuo Okubo (CIBR), Joji Tsunada (CIBR), and Takahiro Shinozaki (Science Tokyo) for their invaluable insights and comments that greatly enhanced the manuscript. Lastly, we thank Dr. Lindsay Bremner for her professional editing assistance. This work was supported by Chinese Institute for Brain Research (CIBR), Beijing.

**AUTHOR CONTRIBUTIONS**

S.O. conceptualized the study and prepared the initial manuscript and figures, integrating substantial feedback from K.O. Both authors collaboratively revised the text and figures and contributed to the finalization of the review manuscript.

**COMPETING INTERESTS**

The authors declare no competing financial interests.

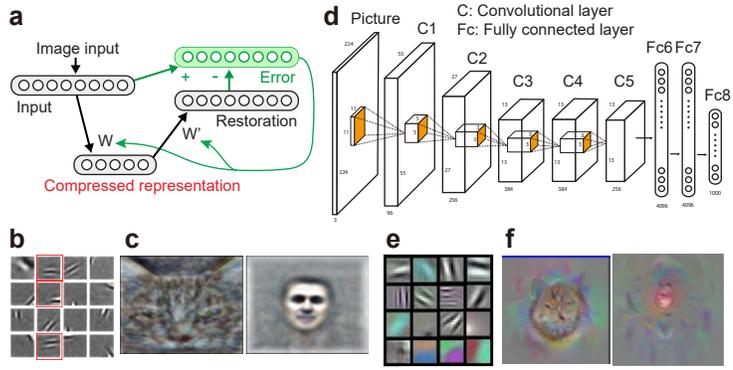

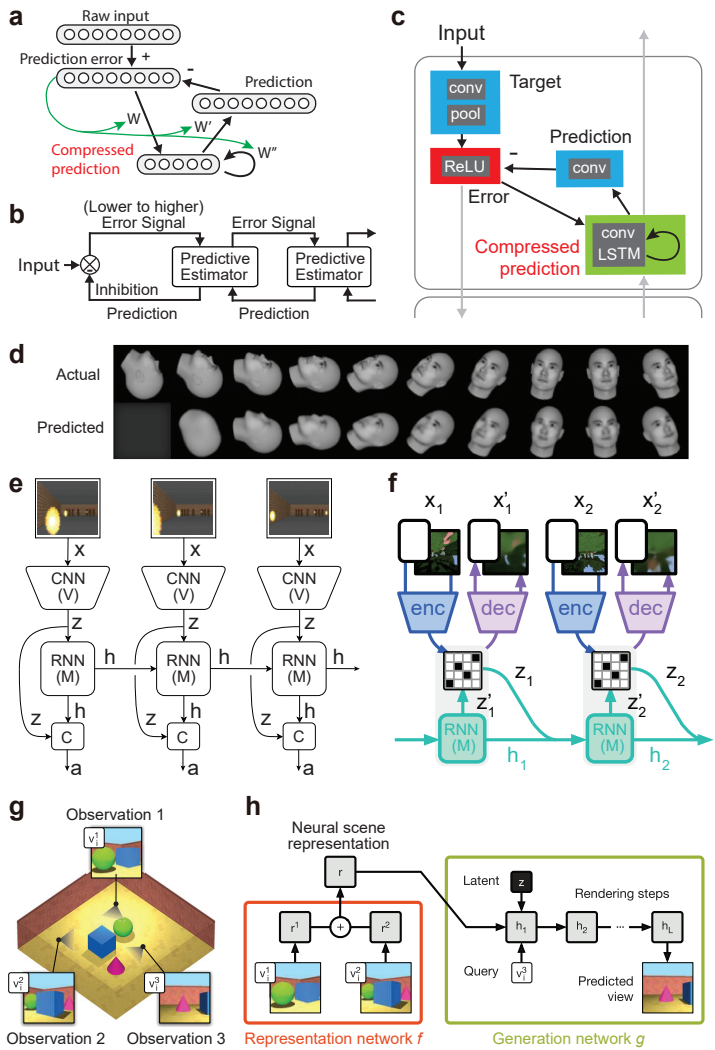

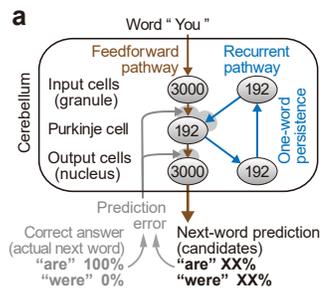 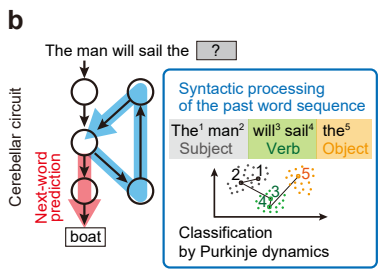

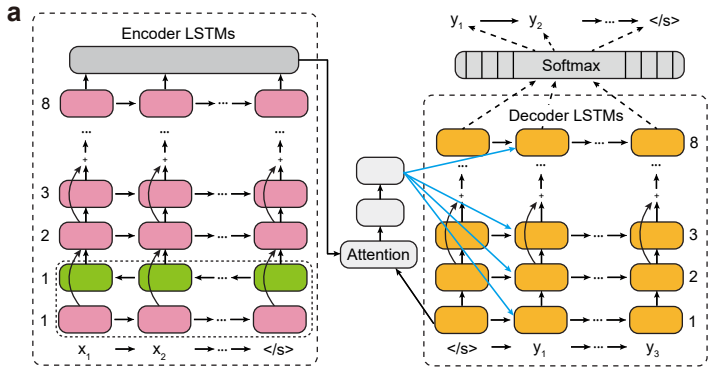

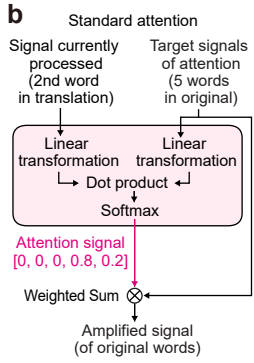
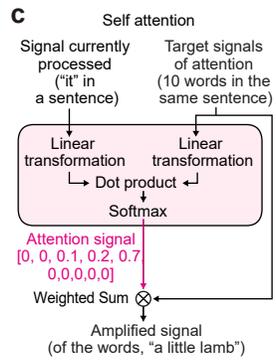
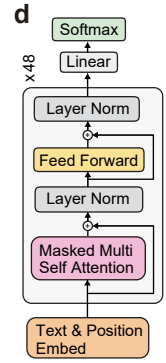

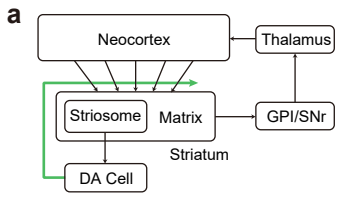
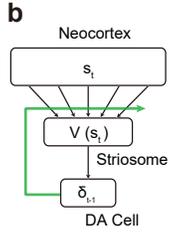
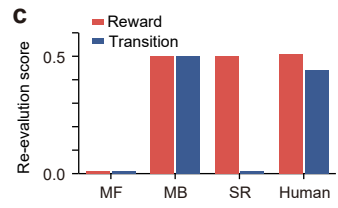
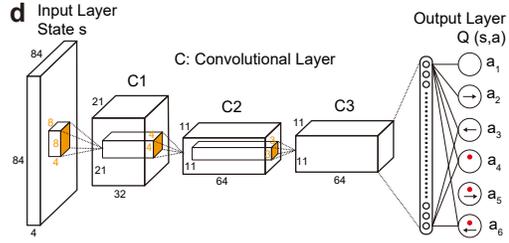

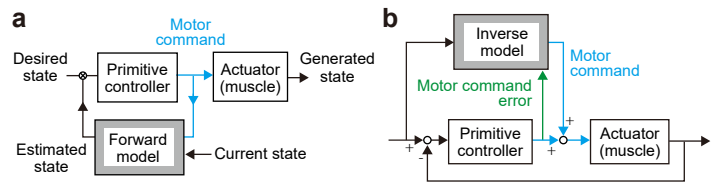

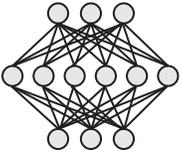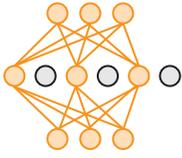

# Supplementary Figures

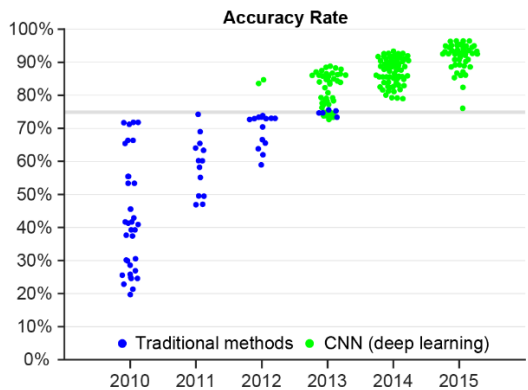

**Supplementary Figure 1.** The dramatic success of CNN in object recognition from visual images. From 2010 to 2012, traditional image recognition methods that did not use neural networks (blue dots; each dot represents the performance of each team in a competition) struggled to surpass the 75% accuracy barrier (thick gray line). In 2012, AlexNet of Hinton group overwhelmingly won the competition with an 85% accuracy rate (green dot, 2012). Since then, CNNs, heavily influenced by AlexNet, have become the dominant approach in image recognition, achieving a 96% accuracy rate by 2015 (ResNet, GoogleNet).

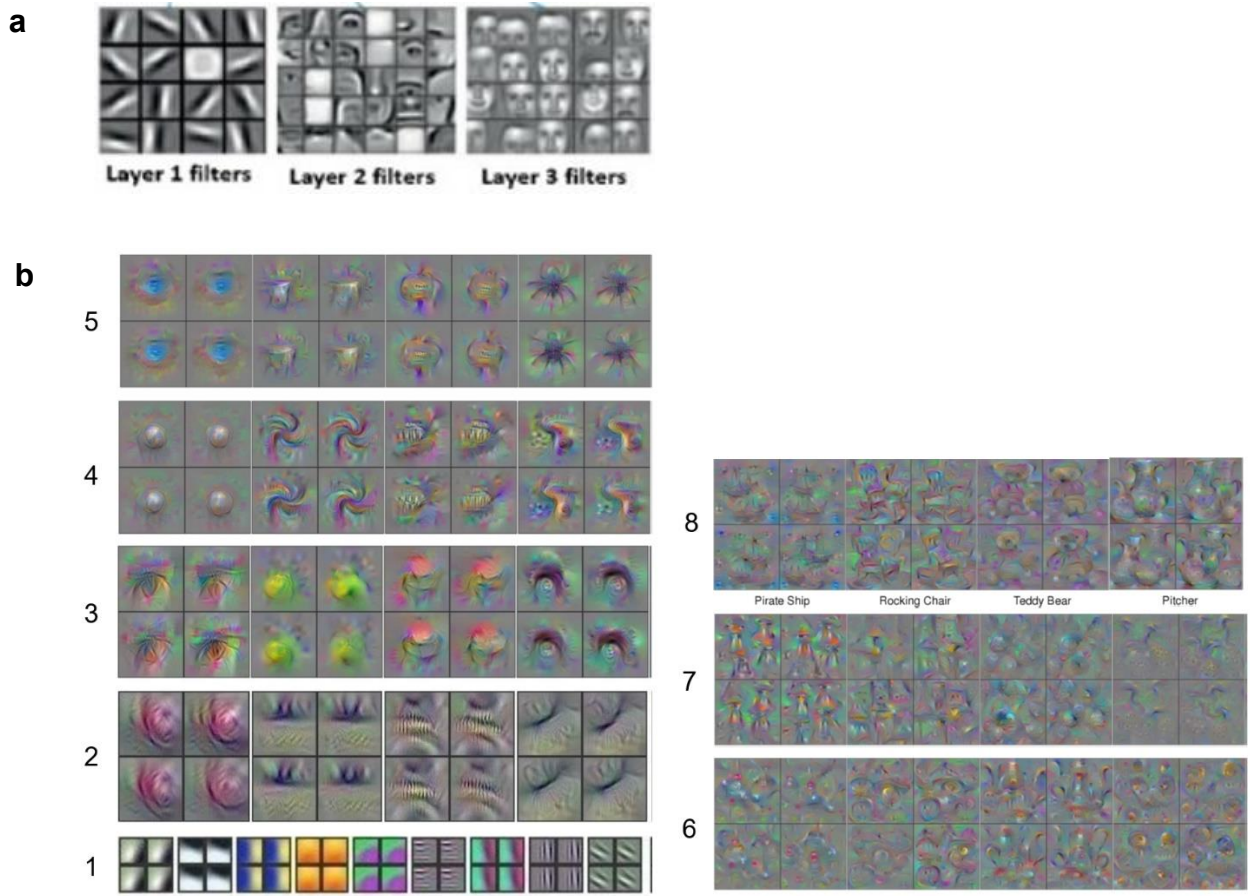

**Supplementary Figure 2.** Comparison of intermediate processing stages between CNNs trained with unsupervised vs supervised learning. **a**. A CNN trained with unsupervised autoencoder learning. Greedy Layer-wise Training was conducted at each layer of a three-layer CNN for natural image input. Intermediate processing was examined by visualizing the optimal stimulus (image input that maximizes the output signal of a particular neuron) for a convolutional neuron at each layer. The first layer responded well to Gabor-function inputs suited for edge detection, the second layer to parts of objects (e.g., parts of a face), and the third layer to the whole images of objects (e.g., a whole human face). Note that the scales of the three diagrams are entirely different. The second layer responds to a smaller visual field than the third layer, and the first layer responds to an even smaller visual field. **b**. Intermediate processing of a CNN (AlexNet) trained with supervised learning. Layers 1-5 are CNN layers (6-8 are fully connected layers), with the first layer responding most strongly to Gabor functions, the subsequent layers responding to parts of objects, and the final CNN layer responding most strongly to the whole objects. The neurons in the 8th layer are responsible for outputting object classifications, responding most strongly to the visual images of the objects they are responsible for (e.g., the neuron responsible for classifying the pitcher responds most strongly to visual images of pitchers). Panel **a** adapted from Wang Y. et al. (2021) (CC BY 4.0). Panel **b** adapted from Yosinski J. et al. (2015) (CC BY-NC-SA 3.0).

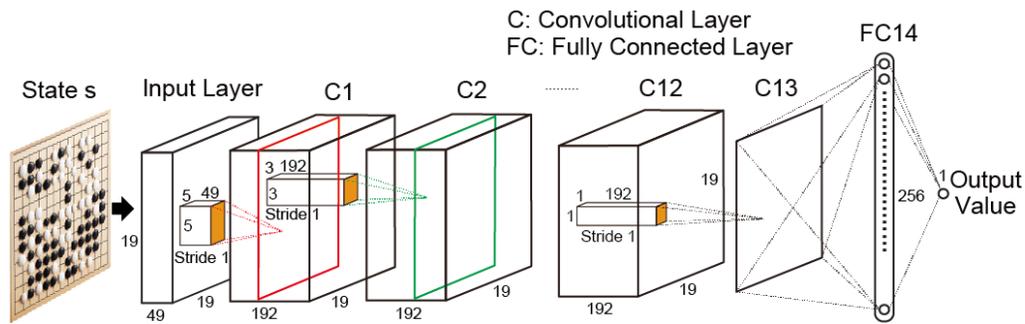

**Supplementary Figure 3**. The design philosophy and network structure of Alpha GO. After DQN (2015), DeepMind developed Alpha GO (2016) and Alpha GO Zero (2017) to conquer GO, a game thought to be impossible to evaluate using reinforcement learning due to its vast search space. DeepMind achieved this by deepening the Deep Q Network and combining it with Monte Carlo Tree Search, which focuses on exploring important paths in the vast search space (tree-shaped search), succeeding in creating Alpha GO (2016) and Alpha GO Zero (2017) that overwhelmed the human world champion. These belong to model-based reinforcement learning machines because they perform "looking ahead" to incorporate the value of future situations (stone patterns; positions) into the value of the current situation. However, Alpha GO and Alpha GO zero maintained the structure similar to that of DQN, with the deep sensory processing part connected to the output part for estimating reinforcement-learning parameters and the entire circuits are updated through backpropagation of the estimation error. Since the transition of the situation (next position) is completely determined by the move that has been made, there is no room (or benefit) to create a causal-relationship model through prediction. Reflecting the values of future situations into the current situation's value can be done with simple arithmetic, so no special circuit is needed (For example, if the current value is 0.5 and the value after a few optimal moves is 0.7, the current value is simply replaced with 0.7 or averaged to 0.6). Therefore, despite being a model-based reinforcement learning circuit, Alpha GO and Alpha GO zero have almost the same structure as a model-free reinforcement learning circuit (such as Deep Q Net). The figure shows the structure of Alpha GO's value generation network $V(s)$. Alpha GO has another next-action-generation network $p(s,a)$ (which outputs the probabilities indicating the next action candidates), which also has almost the same structure. Deep Q Net generates the (estimated) action value $Q(s,a)$ to decide the next action (in the form of Q Learning in reinforcement learning), whereas Alpha GO adopts a policy-based approach in reinforcement learning, using the (estimated) current value $V(s)$ and the next action generator $p(s,a)$ (the set of the probabilities for the next-action candidates is called a "policy" in reinforcement learning). The common structured parts between the value network, $V(s)$, and the next-action-generation network, $p(s,a)$, were integrated into a single shared circuit in Alpha GO Zero. Figure adapted with permission from Kitazawa S. (2020).